\documentclass{aa}

\usepackage{txfonts}
\usepackage{graphicx}
\usepackage[english]{babel}
\usepackage{natbib}
\usepackage[pdftex]{hyperref}
\bibpunct{(}{)}{;}{a}{}{,}

\begin{document}

\title{The evolution of ultracompact X-ray binaries}

\author{
L.~M.~van~Haaften \inst{\ref{radboud}} \and
G.~Nelemans \inst{\ref{radboud},\ref{leuven}} \and
R.~Voss \inst{\ref{radboud}} \and
M.~A.~Wood \inst{\ref{fit}} \and
J.~Kuijpers \inst{\ref{radboud}}
}

\institute{
Department of Astrophysics/ IMAPP, Radboud University Nijmegen, P.O. Box 9010, 6500 GL Nijmegen, The Netherlands, \email{L.vanHaaften@astro.ru.nl} \label{radboud} \and
Institute for Astronomy, K.U. Leuven, Celestijnenlaan 200D, 3001 Leuven, Belgium \label{leuven} \and
Department of Physics and Space Sciences, Florida Institute of Technology, 150 W. University Blvd., Melbourne, FL 32901, USA \label{fit}
}


\abstract{Ultracompact X-ray binaries (UCXBs) typically consist of a white dwarf donor and a neutron star or black hole accretor. The evolution of UCXBs and very low mass ratio binaries in general is poorly understood. In particular, the dynamical behavior of an accretion disk extending to a large radius (relative to the orbit) is unclear.}
{We investigate the evolution of UCXBs in order to learn for which mass ratios and accretor types these systems can exist, and if they do, what are their orbital and neutron star spin periods, mass transfer rates and evolutionary timescales.}
{We compute tracks of a binary containing a Roche-lobe overflowing helium white dwarf in which mass transfer is driven by gravitational wave emission. For different assumptions concerning accretion disk behavior we calculate for which system parameters dynamical instability, thermal-viscous disk instability or the propeller effect emerge. The significance of these processes during the evolution of an UCXB is considered.}
{At the onset of mass transfer, the survival of the UCXB is determined by how efficiently the accretor can eject matter in the case of a super-Eddington mass transfer rate. At later times, the evolution of systems strongly depends on the binary's capacity to return angular momentum from the disk to the orbit. We find that this feedback mechanism most likely remains effective even at very low mass ratio. In the case of steady mass transfer, the propeller effect can stop accretion onto recycled neutron stars completely at a sufficiently low mass transfer rate, based on energy considerations. However, mass transfer will likely be non-steady because disk instability allows for accretion of some of the transferred matter. Together, the propeller effect and disk instability cause the low mass ratio UCXBs to be visible a small fraction of the time at most, thereby explaining the lack of observations of such systems.}
{Most likely UCXBs avoid late-time dynamically unstable mass loss from the donor and continue to evolve as the age of the Universe allows. This implies the existence of a large population of low mass ratio binaries with orbital periods $\sim 70 - 80$ min, unless some other mechanism has destroyed these binaries. Even though none have been discovered yet, black hole UCXBs could also exist, at orbital periods of typically $100 - 110$ min.}

\authorrunning{van~Haaften et al.}

\maketitle

\section{Introduction}

Ultracompact X-ray binaries (UCXBs) are a subclass of low-mass X-ray binaries and consist of a white dwarf or helium star losing mass to a neutron star or black hole, at a sub-hour orbital period \citep{savonije1986}. The short orbital periods point to white dwarf or helium burning star (sdB star) donors, as those are the only known stellar types that have the same size as the donor Roche lobe corresponding to these orbital periods \citep{nelson1986}. The observed helium and carbon-oxygen composition of the transferred matter confirms this picture \citep{schulz2001,nelemans2004,nelemans2006}. The accretor type can be identified via its inferred mass, magnetic field, spin period, explosive nuclear fusion on its surface or inner accretion disk behavior, and can be either a neutron star or a black hole (the latter have not yet been observed in UCXBs). To date about $30$ UCXBs and candidates have been identified \citep{zand2007}.

UCXBs are important objects to study because of the absence of hydrogen in the accretion disk and X-ray bursts \citep{zand2005}, their gravitational wave signal \citep{nelemans2009}, and them being tests for the common-envelope phase \citep{nelemans2010b}. Furthermore, UCXBs are candidate progenitors of radio millisecond pulsars \citep{alpar1982} because of the high number of accreting millisecond pulsars found in UCXBs \citep{wijnands2010}. The ratio between the number of black hole and neutron star accretors constrains the high-mass end of the binary initial mass function.

\subsection{Evolutionary history}

The formation of ultracompact X-ray binaries starting from a zero-age main sequence binary involves one or two common envelope stages and a supernova. The initially more massive star evolves off the main sequence and starts losing mass to its companion. Depending on the mass ratio and stage of donor evolution, this may happen in an unstable way, leading to a common envelope. The orbit decays significantly due to energy and angular momentum loss via friction. If the system does not merge, it emerges as a much shorter-period system consisting of the core of the giant and a rejuvenated main sequence companion. There are several scenarios for the subsequent evolution; the core of the giant may develop into a core-collapse supernova, or the main-sequence companion may evolve into a (sub)giant, causing a second common envelope before it becomes a supernova itself \citep{tutukov1993,podsiadlowski2002}. If precisely one of the stars becomes a supernova and the system is not unbound during this event, the system can develop into a neutron star/black hole - white dwarf/helium star binary. If neither of the components is massive enough to become a core-collapse supernova, a second common envelope can lead to a double white dwarf or a white dwarf - helium star binary. An accretion-induced collapse of a white dwarf may still yield an UCXB, once mass transfer has resumed \citep{vandenheuvel1984}.

More models for UCXB formation exist. In globular clusters, home to about one-third of the presently known UCXB population, formation may be dominated by dynamical interactions \citep{verbunt1987,ivanova2005,voss2007}.
After a system with a neutron star or black hole component has formed, UCXBs can evolve through $3$ scenarios: via a white dwarf donor \citep{yungelson2002}, helium-star donor \citep{savonije1986,yungelson2008} or a main-sequence donor \citep{podsiadlowski2002,sluys2005a,ma2009}. However, after a relatively short time of mass transfer, each donor becomes degenerate and the subsequent evolution is very similar for each scenario. Therefore, we consider only the white dwarf donor scenario.

As suggested by the rarity of UCXBs, the initial stellar and binary parameters must be finely tuned in order to arrive at a stable ultracompact configuration. Several things can go wrong: binaries can either evolve too slowly to have two post-main sequence components, be too wide to have interacting components, merge during a common envelope, possibly explode as a single degenerate type Ia supernova, become unbound by a supernova, or experience two supernovae.

\subsection{Present research}

The objective of this paper is to investigate the occurrence of a dynamical instability in low mass ratio UCXBs, and the influence on UCXB evolution of magnetosphere-disk interactions and the thermal-viscous disk instability.
By improving our understanding of all evolutionary stages of UCXBs, we will be able to better estimate their observational properties, such as X-ray luminosity and orbital period, fraction of time visible and lifetime. In a forthcoming paper, we will use the results to predict the present-day UCXB population, which can be compared to observations by the Galactic Bulge Survey \citep{jonker2011}.

In particular the behavior of old, low mass ratio ($q < 0.01$) and low mass transfer rate systems is unclear. None have been discovered, even though their long evolutionary timescales suggest that many should exist if no disruptive process emerges at some point. Are they invisible to our instruments (most of the time), or do they no longer exist?

Section \ref{overview} gives an overview of regular UCXB evolution and the main angular momentum flows in the absence of instabilities. Several complications that may play an important role during the evolution of an UCXB are mentioned in Sect. \ref{complic}. In the method, Sect. \ref{method}, we describe the evolutionary tracks as a function of the degree of feedback of angular momentum from accretion disk to orbit, the propeller effect and the disk instability model. In the results, Sect. \ref{results}, we analyze the tracks, consider how likely limited feedback is, and look into the magnetic field, both for constant mass transfer and non-steady behavior such as outbursts. In Sect. \ref{disc} we discuss the results and give a conclusion. The appendix contains fits of evolutionary tracks and some analytical approximations.

\section{Overview of UCXB orbital mechanics}
\label{overview}

The evolution of a binary consisting of only degenerate stars or black holes cannot be driven by single star evolution, since such components hardly evolve, if at all, and certainly do not increase their radius in the absence of interaction with other stars. Since we assume white dwarfs to be cold, i.e. thermally relaxed, angular momentum loss and redistribution are the only processes that can alter the system.

\subsection{Angular momentum flows}
\label{amflows}

The angular momentum flows in an UCXB are given by

\begin{equation}
    \label{j_balance}
    \dot{J}_\mathrm{orb} = \dot{J}_\mathrm{\textsc{gwr}} + \dot{J}_\mathrm{stream} + \dot{J}_\mathrm{torque} + \dot{J}_\mathrm{eject},
\end{equation}
where $\dot{J}_\mathrm{orb} < 0$ is the change in orbital angular momentum, $\dot{J}_\mathrm{\textsc{gwr}} < 0$ the loss to gravitational wave emission, $\dot{J}_\mathrm{stream} < 0$ angular momentum advected from the donor to the accretion disk along with the transferred matter, $\dot{J}_\mathrm{torque} > 0$ is the return of angular momentum from the disk to the orbit by means of a tidal torque between the outer disk and the donor \citep{lin1979,frank2002book} and $\dot{J}_\mathrm{eject} < 0$ is the angular momentum carried by matter ejected from the system. The high surface gravity of degenerate stars prohibits significant wind mass loss from the donor, so spin angular momentum loss via magnetic braking can be neglected. Change in accretor spin can present a sink or source of angular momentum \citep{marsh2004}, but contrary to white dwarf accretors, this is negligible compared to the angular momentum transport to and from the disk \citep{priedhorsky1988} for neutron star and black hole accretors. 

When the system has an accretion disk, $\dot{J}_\mathrm{stream}$ is precisely balanced by the feedback flow in the opposite direction, $\dot{J}_\mathrm{torque}$ \citep{priedhorsky1988}. In that case, the accretion disk does not gain or lose angular momentum, which leaves orbital angular momentum loss via gravitational wave radiation as the sole process driving binary evolution, giving the usual expression

\begin{equation}
    \label{j_balance2}
    \dot{J}_\mathrm{orb} = \dot{J}_\mathrm{\textsc{gwr}} + \dot{J}_\mathrm{eject}.
\end{equation}
However, below we will consider the more general case. The orbital period of two stars must be rather short for gravitational wave emission to be able to significantly shrink the orbit within the age of the Universe, e.g. shorter than $14$ hr in the case of two $1\ M_{\odot}$ components. Once the orbit has shrunk to the point where the Roche lobe of a white dwarf component is smaller than the associated star's volume, the white dwarf starts to lose surface matter.

\subsection{Stable mass transfer}
\label{appmt}

\begin{figure}
\resizebox{\hsize}{!}{\includegraphics{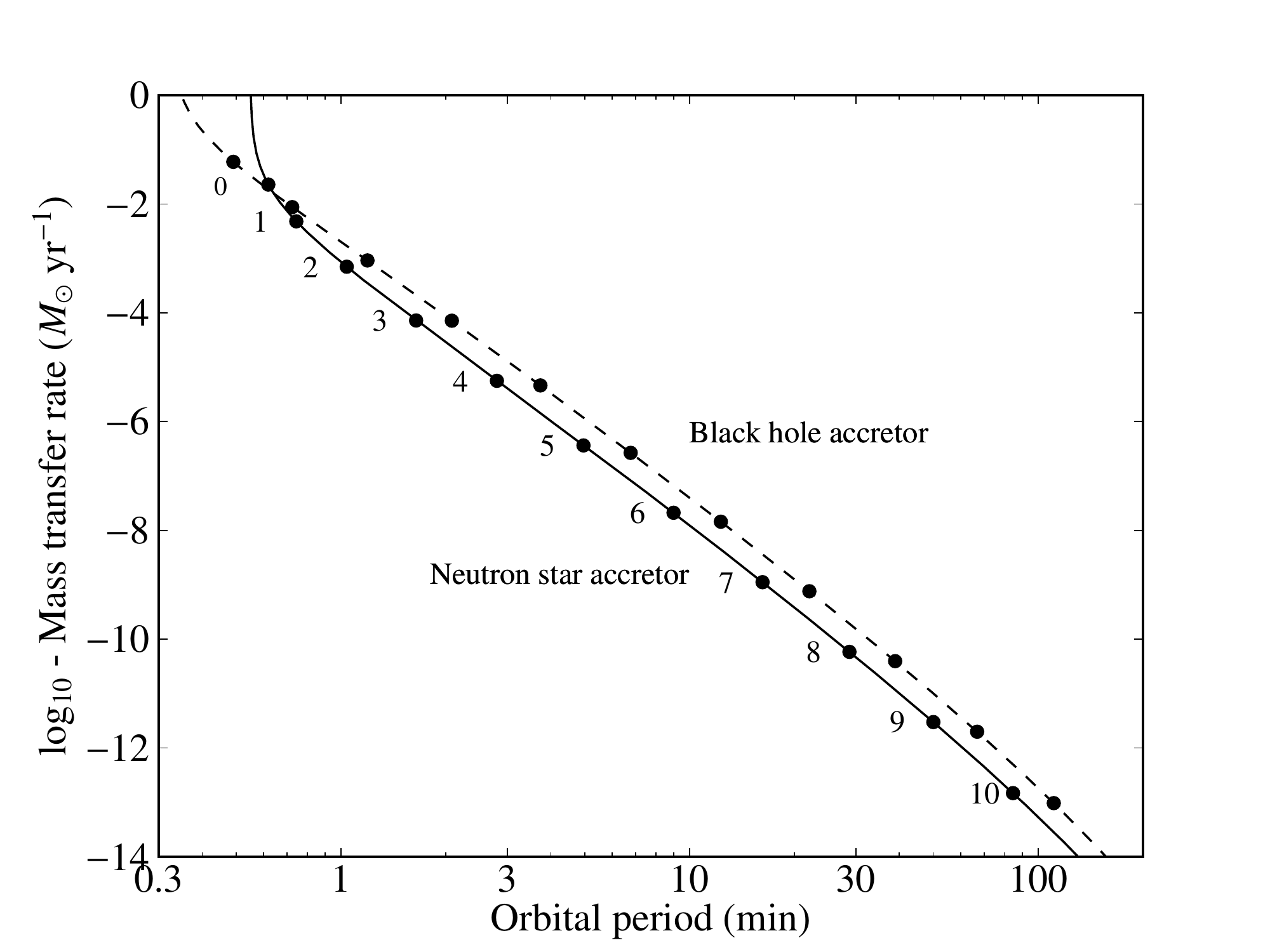}} 
\caption{Tracks for a helium white dwarf donor with an initially $1.4\ M_{\odot}$, $12\ \mbox{km}$ radius neutron star (solid) and $10\ M_{\odot}$ black hole (dashed) accretor. The tracks start at the respective dynamical instability limits of $0.83\ M_{\odot}$ and $1.04\ M_{\odot}$, since systems can only survive if they start with a lower donor mass (Sect. \ref{dscrit}). The isotropic re-emission limit is ignored in the first part of the track. The numbers correspond to the two circles on their right-hand side and represent the logarithm of the system age (\mbox{yr}), which is the time since the onset of mass transfer. The relative location of both tracks will be discussed in Sect. \ref{macc}.}
\label{fig:wdtracks}
\end{figure}

To calculate the mass transfer rate, we assume that the donor precisely fills its Roche lobe at all times $t$ during stable mass loss (in reality the donor slightly overfills its Roche lobe, but this distance is small compared to the Roche-lobe radius and can be neglected):

\begin{equation}
    \label{basic}
    R_\mathrm{d}(t) = R_\mathrm{L}(t) \qquad \mathrm{and} \qquad \dot{R}_\mathrm{d}(t) = \dot{R}_\mathrm{L}(t) \qquad \forall\ t
\end{equation}
where $R_\mathrm{d}$ is the donor radius and $R_\mathrm{L}$ is the effective radius of the donor Roche lobe. Equation \ref{basic} implies that the ratio of the component masses and the donor radius determines the semi-major axis and orbital period, which increase with time. The change in $R_\mathrm{d}$ and $R_\mathrm{L}$ can be separated in a mass-transfer dependent and -independent part as \citep{soberman1997}

\begin{eqnarray}
    \label{rzeta}
    \frac{\dot{R}_\mathrm{d}}{R_\mathrm{d}} &=& \left(\frac{\dot{R}_\mathrm{d}}{R_\mathrm{d}} \right)_\mathrm{evo} + \zeta_\mathrm{d}\ \frac{\dot{M}_\mathrm{d}}{M_\mathrm{d}} \\ 
    \label{rzeta2}
    \frac{\dot{R}_\mathrm{L}}{R_\mathrm{L}} &=& \left(\frac{\dot{R}_\mathrm{L}}{R_\mathrm{L}} \right)_\mathrm{aml} + \zeta_\mathrm{L}\ \frac{\dot{M}_\mathrm{d}}{M_\mathrm{d}} 
\end{eqnarray}
where $\zeta_\mathrm{i} \equiv \partial \ln R_\mathrm{i}/\partial \ln M_\mathrm{d}$ \citep{hut1984} represents the change in $R_\mathrm{i}$ resulting from mass transfer or loss, i.e. a change in donor mass $M_\mathrm{d}$ ($R_\mathrm{i} \propto M_\mathrm{d}^{\zeta_\mathrm{i}}$). The subscript \emph{evo} denotes single star evolution and \emph{aml} denotes angular momentum loss from the system (in the absence of mass transfer and loss). The advantage of the above separation of the change in radii is that it allows for more easily solving the equilibrium mass loss rate.
From Roche lobe geometry and the orbital angular momentum equation follows that $(\partial \ln R_\mathrm{L})_\mathrm{aml} = (\partial \ln a)_\mathrm{aml} = 2 (\partial \ln J_\mathrm{orb})_\mathrm{aml}$ in the case of fixed component masses ($a$ is the semi-major axis). Combining the above and taking ($\partial \ln R_\mathrm{d})_\mathrm{evo} = 0$ for the fully cooled, non-evolving, white dwarf case, leads to the (negative) donor mass loss rate\footnote{An approximate analytic solution of Eq. (\ref{mt2}) is given in Appendix \ref{appa}, along with fitted tracks in Appendix \ref{wdfits}.}

\begin{equation}
    \label{mt2}
    \frac{\dot{M}_\mathrm{d}}{M_\mathrm{d}} = \frac{2}{\zeta_\mathrm{d}-\zeta_\mathrm{L}} \cdot \left( \frac{\dot{J}}{J} \right)_\mathrm{\textsc{gwr}}
\end{equation}
where orbital angular momentum (unrelated to mass transfer or mass loss) is lost exclusively by gravitational wave emission (Eq. \ref{j_balance2}), given by \citep{landau1975}. The effect of angular momentum loss via mass ejection is contained in $\zeta_\mathrm{L}$ and will be discussed in Sect. \ref{rlrad}. In Fig. \ref{fig:wdtracks} we show the mass transfer rate given by Eq. (\ref{mt2}) for $\zeta_\mathrm{d}$ given in Sect. \ref{wdrad}.

Figure \ref{fig:wdtracks} shows how the mass transfer rate decreases with increasing orbital period and time. The white dwarf grows in size as it loses mass, and since it is assumed to fill its Roche lobe at all times (Eq. \ref{basic}), the orbital separation and hence orbital period must increase as well. Since gravitational wave radiation becomes significantly weaker in a wider orbit, the mass transfer rate decreases. This can be understood as follows: mass transfer from a less massive to a more massive component in itself tends to widen the orbit, and the weaker the orbit-shrinking gravitational waves, the less mass transfer is needed to keep the orbit wide enough to accommodate the donor.

If the binary transfers too much matter, the donor detaches and mass transfer stops. Conversely, if too little matter is transferred, the donor overfills its Roche lobe more and more, increasing mass loss. An equilibrium is reached; the system naturally arrives at a mass transfer rate for which the donor overfills its Roche lobe by the right amount.

\subsection{Dynamical stability}
\label{dscrit}

From Eq. (\ref{mt2}) it can be seen that the mass transfer rate becomes unbounded when $\zeta_\mathrm{L}$ and $\zeta_\mathrm{d}$ approach each other. Dynamical stability of the donor requires that, upon mass loss, the change in $R_\mathrm{L}$ exceeds the change in $R_\mathrm{d}$, i.e.\footnote{Note that this criterion has a different meaning than the second part of Eq. (\ref{basic}). Here, change in radii resulting from only mass loss is considered, whereas Eq. (\ref{basic}) applies when gravitational wave radiation is included as well.}

\begin{equation}
    \dot{R}_\mathrm{L} > \dot{R}_\mathrm{d}.
\end{equation}
Dividing both sides by $R_\mathrm{d} = R_\mathrm{L}$ (Eq. \ref{basic}) and by $\dot{M}_\mathrm{d}/M_\mathrm{d}< 0$ yields the well known criterion (e.g. \citet{webbink1985,pols1994})\footnote{A stability criterion expressed in $\zeta$-values, i.e. logarithmic derivatives, is to be preferred over linear derivatives of radius to donor mass because the expressions for $\zeta_\mathrm{L}$ and $\zeta_\mathrm{d}$, Eqs. (\ref{zetarl}) and (\ref{zetawd}) respectively, are simpler than the expression for $\partial R_\mathrm{i}/\partial M_\mathrm{d} = \zeta_\mathrm{i} \cdot R_\mathrm{i}/M_\mathrm{d}$ as the factors $R_\mathrm{i}$ and $M_\mathrm{d}$ introduce several extra variables.}

\begin{equation}
    \label{stab}
    \zeta_\mathrm{L} < \zeta_\mathrm{d}.
\end{equation}
Physically this means that when a donor that does not obey this criterion starts transferring matter via Roche-lobe overflow, its volume grows faster than the volume of the Roche lobe it is contained in. The donor will overflow its Roche lobe more and more, leading to runaway mass loss on the dynamical timescale of the donor ($\sim 5$ s for an $0.6\ M_{\odot}$ white dwarf), and followed by disruption of the donor.

\begin{figure}
\resizebox{\hsize}{!}{\includegraphics{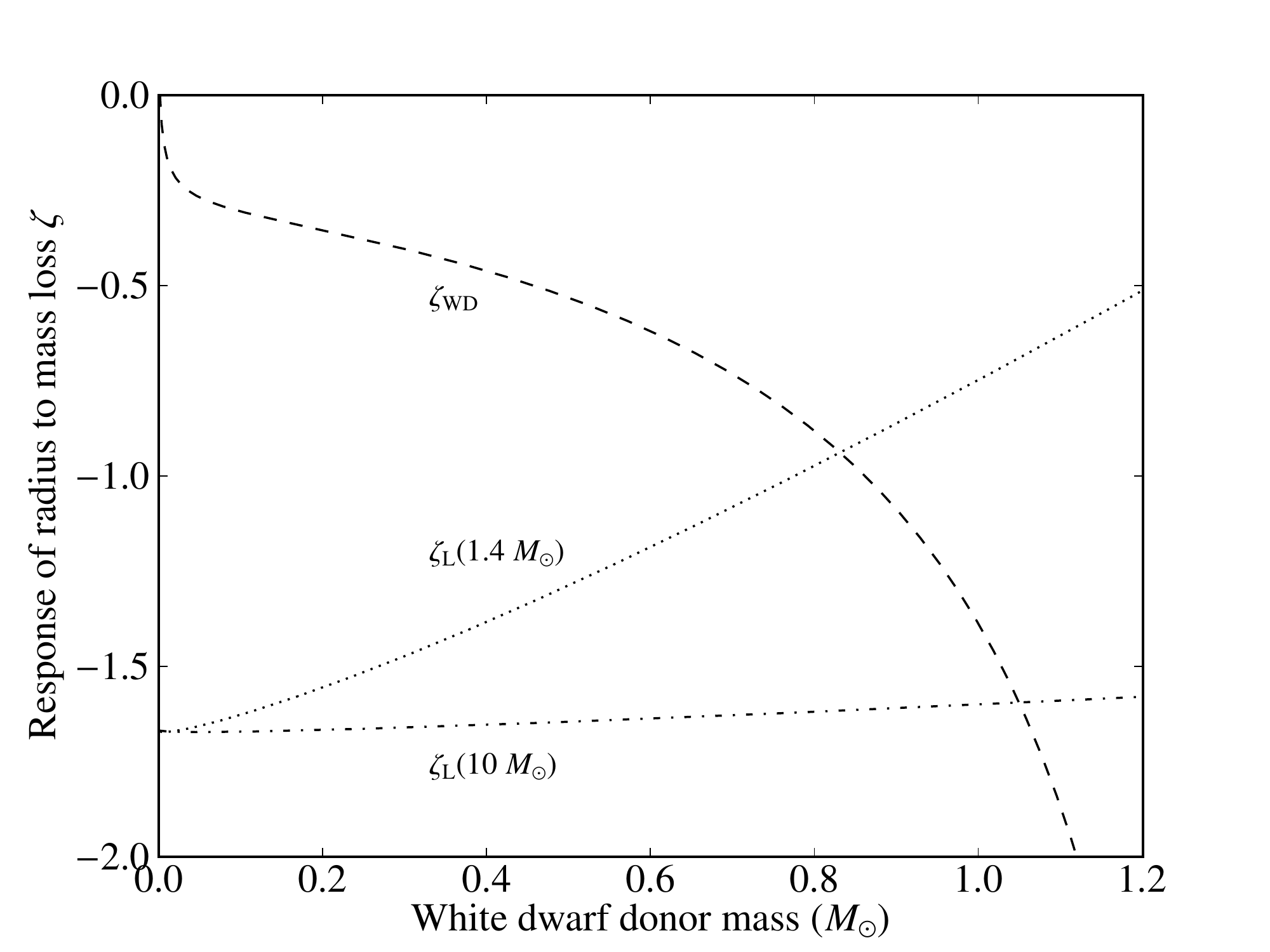}} 
\caption{Mass-radius relation exponent $\zeta_\mathrm{L}$ (Sect. \ref{rlrad}) for a $1.4\ M_{\odot}$ accretor (dotted) and a $10\ M_{\odot}$ accretor (dash-dotted) and the zero-temperature white dwarf $\zeta_\mathrm{d}$ (dashed, Sect. \ref{wdrad}). Nearly all matter is assumed to be lost via the isotropic re-emission mechanism (Sect. \ref{iresect}).}
\label{fig:zetas_simple}
\end{figure}

Near the dynamical instability limit, the mass transfer rate becomes in principle arbitrarily high, as shown by the upper left part of the tracks in Fig. \ref{fig:wdtracks}. This means that in order to find the critical mass ratio at which this instability occurs, we have to consider $\zeta_\mathrm{L}$ for the case in which a fraction $1$ of transferred mass is ejected. Figure \ref{fig:zetas_simple} shows that for a system with an accretor mass $M_\mathrm{a} = 1.4\ M_{\odot}$ (dotted curve), mass transfer is dynamically unstable if $M_\mathrm{d} > 0.83\ M_{\odot}$. In the case of a $10\ M_{\odot}$ accretor (dash-dotted) this becomes $M_\mathrm{d} > 1.04\ M_{\odot}$.

\subsection{Isotropic re-emission}
\label{iresect}

Even when the donor mass is low enough to avoid a dynamical instability, the high amount of mass it needs to lose to avoid a merger of the binary components shortly after the onset of mass transfer \citep{vandenheuvel1984bons} may be too much for the accretor to accrete or eject. The ultimate condition for survival is how efficiently the accretor can eject matter in the case of a super-Eddington mass transfer rate. The upper limit of mass that can be removed from the system follows from the energy balance. By assuming that precisely the Eddington limit is accreted, the large amount of gravitational energy liberated can in principle be employed to unbind additional transferred matter (with energy corresponding to the first Lagrangian point) from the system. This mechanism is called isotropic re-emission. \citet{soberman1997} and \citet{tauris1999} described the mechanism as matter transferred via an accretion disk to the vicinity of the accretor, from where it is subsequently ejected as a fast, isotropic wind.

\subsection{The deciding stability criterion at the onset of mass transfer}
\label{deciding}

\begin{figure}
\resizebox{\hsize}{!}{\includegraphics{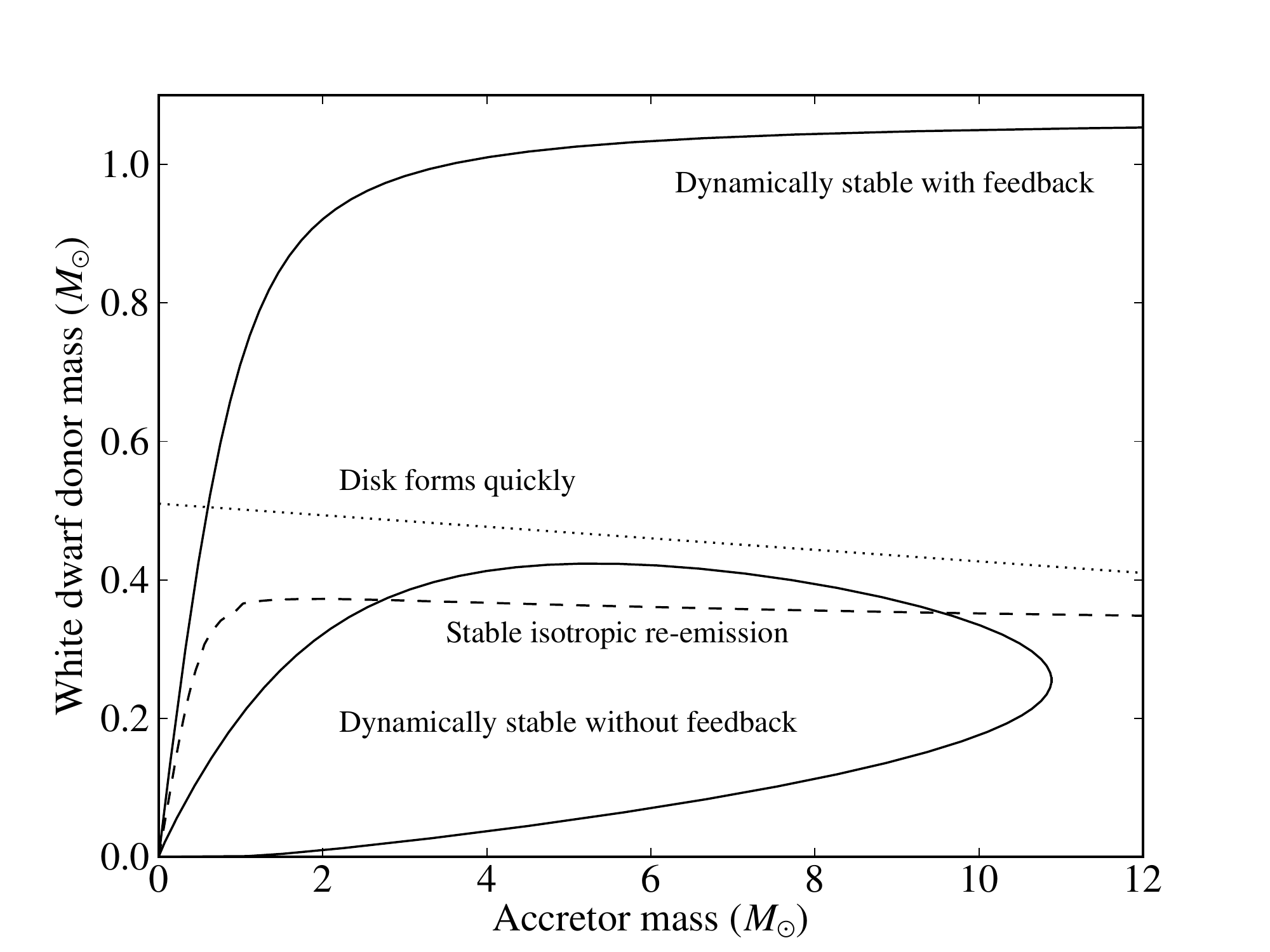}} 
\caption{Stability regimes in terms of zero-temperature helium white dwarf donor mass versus accretor mass. The upper solid curve is the upper limit for dynamical stability in the presence of an accretion disk that feeds back angular momentum, where all transferred matter is lost via isotropic re-emission. The dashed curve is the upper limit for isotropic re-emission (i.e. if the system lies above this curve, the accretor will be engulfed by transferred matter). The lower solid curve encloses the parameter space where systems are dynamically stable in the absence of angular momentum feedback. The dotted curve indicates a very rough estimate of the upper donor mass for which there is enough time to form an accretion disk and suppress the no-feedback instability, based on \citet{verbunt1988}.}
\label{fig:crith}
\end{figure}

The onset of Roche-lobe overflow is a very important event in the evolution of a binary. White dwarf donors that are too massive lose mass in a dynamically unstable way (Sect. \ref{dscrit}) and the systems containing these do not survive. Figure \ref{fig:crith} (upper solid curve) shows that systems with a $1.4\ M_{\odot}$ neutron star accretor are dynamically unstable if the white dwarf donor is more massive than $0.83\ M_{\odot}$. Since there is no accretion disk present yet when mass transfer first starts, there is no tidal torque transferring back angular momentum to the orbit (to be discussed in Sect. \ref{explain_feedback}), and the criterion becomes stricter; the system must lie within the lower solid curve in Fig. \ref{fig:crith}. However, if the donor is less massive than $\sim 0.50\ M_{\odot}$ (dotted curve), the disk forms on a timescale short enough to start transferring back enough angular momentum to stop the dynamical instability \citep{verbunt1988}.
Moreover, the $0.50\ M_{\odot}$ disk formation criterion itself is replaced by an even stricter criterion: the isotropic re-emission limit of $M_\mathrm{d} \approx 0.37\ M_{\odot}$ (dashed curve).\footnote{\citet{yungelson2002} found a value of $0.44\ M_{\odot}$, for a $1.433\ M_{\odot}$ neutron star accretor, due to a different adopted isotropic re-emission prescription combined with the relatively weak sensitivity to donor mass of both mass transfer rate and isotropic re-emission rate.
} In this paper we assume that isotropic re-emission is the criterion which in practice determines whether a system survives the onset of mass transfer, or merges.\footnote{If isotropic re-emission would fail to prevent a merger, UCXB could only be formed via an initially non-degenerate donor like a helium star, with the constraint that it should not become fully degenerate before reaching a mass below $0.08\ M_{\odot}$, when mass transfer becomes sub-Eddington.}

If the binary system survives the onset of mass transfer, it will settle in a state of stable mass transfer, as described in Sect. \ref{appmt}.

\subsection{Complications}
\label{complic}

The perpetually expanding binary with declining mass transfer rate described above is idealized. In reality, several processes can potentially disturb continuous accretion or even disrupt the binary, and they may become important in view of the low mass ratios and mass transfer rates expected for compact systems driven by gravitational wave radiation.

\subsubsection{Limited feedback of angular momentum at low mass ratio}
\label{explain_feedback}

In Sect. \ref{dscrit} the possibility of a dynamical instability at the onset of mass transfer was discussed. According to Fig. \ref{fig:zetas_simple} such an instability can only occur at a high donor mass of $M_\mathrm{d} \gtrsim 0.8\ M_{\odot}$, depending on accretor mass. This is no longer necessarily true if an extra angular momentum sink is introduced. Here we look into possible reasons for limited feedback of angular momentum from disk to orbit.

Angular momentum is advected from the donor to the accretion disk along with transferred matter. Initially, when a disk has yet to form, matter accumulates in a ring around the accretor and, due to viscous friction, spreads out radially to form an accretion disk. The disk keeps expanding until it reaches a radius that is so large that all added angular momentum can be returned to the orbit via the tidal torque between the donor and the outer disk, and no further \citep[see][chap.~5]{frank2002book}. A torque exists because the disk is slightly elongated due to the tidal force of the donor, and also slightly ahead of the donor because of the shorter orbital period and a delay in adjusting its shape. At the same time, the outwards transport of angular momentum through the disk allows for matter to move inwards, and eventually to be accreted onto the central object.

Possible causes for reduced feedback include a viscosity that is too low and tidal torque that is too weak \citep{lin1979,ruderman1983,hut1984}, and the resumption of mass transfer after a detached stage caused by orbital widening, which in turn is due to mass loss caused by an accretion-induced collapse of a white dwarf accretor \citep{vandenheuvel1984}.
Recently, \citet{yungelson2006,lasota2007}
stated that it remains unknown what happens to mass transfer and the accretion disk in systems with a mass ratio $q = M_\mathrm{d}/M_\mathrm{a} \lesssim 0.01$. When $q \lesssim 0.02$, the circularization radius exceeds the estimates of the outer radius by \citet{paczynski1977} and \citet{papaloizou1977}, causing the disk to be truncated, and matter to circularize onto unstable orbits, thereby preventing accretion \citep{yungelson2006,lasota2008}.
The donor may cause gaps in the disk at certain radii via destabilizing orbital resonances, causing disk particles to be confined to certain radius ranges. This process is similar to the formation of gaps in the ring system of \object{Saturn} by \object{Mimas} and other moons \citep{franklin1970}, and the Kirkwood gaps in the asteroid belt due to \object{Jupiter}'s gravitational influence. Matter may be unable to pass these gaps, which would prevent both the outward expansion of the disk to the radius required for full angular momentum feedback to the orbit, as well as the corresponding accretion on the inside.

\subsubsection{Occurrence of a dynamical instability at low mass ratio}
\label{occ_inst}

A dynamical instability at low mass ratio may happen only if feedback would be reduced for any of the reasons given in Sect. \ref{explain_feedback}. In the absence of feedback $\dot{J}_\mathrm{torque} =0$ and Eq. (\ref{j_balance}) becomes
\begin{equation}
    \label{j_balance3}
    \dot{J}_\mathrm{orb} = \dot{J}_\mathrm{\textsc{gwr}} + \dot{J}_\mathrm{stream} + \dot{J}_\mathrm{eject},
\end{equation}
The disk would turn into an extra angular momentum sink and the orbit loses more angular momentum than in the case of feedback, and has a stronger tendency to shrink. More mass must be transferred to compensate for this and keep the donor within its Roche lobe, and thereby a new equilibrium is established. For systems with a low mass ratio this effect is stronger because of the geometry of the system: mass entering the disk carries a high specific angular momentum because of the large distance (relative to the semi-major axis) of the first Lagrangian point to the center of mass. At very low mass ratio, the orbit loses so much angular momentum to the disk that no amount of mass transfer can compensate for this; the donor will be disrupted.\footnote{The dynamical instability for \emph{high mass} white dwarf donors on the other hand is a result of the steep mass-radius relation of the donor, and also because the orbit expands less easily when the mass ratio is closer to $1$ (Fig. \ref{fig:zetas_simple}).}

\citet{ruderman1983} were the first to identify this instability, and \citet{ruderman1985} showed that in the case of a helium white dwarf donor, the dynamical instability can be reached after $9\ \mbox{Gyr}$, with $M_\mathrm{d} = 4 \cdot 10^{-3}\ M_{\odot}$ for $M_\mathrm{a} = 1.4\ M_{\odot}$. \citet{bonsema1985} and \citet{verbunt1988} came to a similar conclusion.

On the other hand, \citet{jeffrey1986} reported that the fast spin period of the single millisecond pulsar \object{PSR 1937+214} contradicts a history in a binary followed by donor disruption, because the spin would have been much slower at the evolutionary stage such a disruption would take place.

\citet{bildsten2002} considered finite-temperature white dwarfs and found that those would reach the instability at higher donor mass than zero-temperature white dwarfs, for instance at $M_\mathrm{d} = 0.01$ in the case of a $10^{6}$ K helium white dwarf donor and a $M_\mathrm{a} = 1.4\ M_{\odot}$ accretor.

The questions we want to answer are, 1) could feedback be reduced at some stage in the evolution of an UCXB, and 2) if so, does this lead to a dynamical instability within the age of the Universe?

\subsubsection{Propeller effect}

More processes can potentially limit accretion.
In the case of a neutron star accretor, accretion may be disturbed by the magnetic field. This is the propeller effect, in which the fast-rotating magnetosphere accelerates transferred matter in the azimuthal direction and stops it from accreting \citep{davidson1973,illarionov1975}.

\subsubsection{Thermal-viscous disk instability}
\label{dim_intro}

At low mass transfer rates, the disk instability model describes a thermal-viscous instability resulting from a relatively large and sudden local increase in opacity and viscosity of the disk material \citep{osaki1974,lasota2001} (note, this disk instability is unrelated to the dynamical instability discussed in Sects. \ref{dscrit} and \ref{occ_inst}). The high viscosity causes a much higher accretion rate, called an outburst. Outbursts are alternated by low-viscosity stages during which the disk builds up again. Stable behavior persists if the entire disk has a rather homogeneous degree of ionization; UCXBs in particular can have a stable disk if the mass transfer rate is sufficiently high to keep the entire disk ionized through X-ray irradiation \citep{zand2007}.

\section{Method}
\label{method}

In this section we describe the details of our treatment of the evolutionary stages of UCXBs. The concepts that will play a role during the evolution of UCXBs are discussed in order of appearance.

\subsection{Isotropic re-emission}
\label{ire}

As discussed in Sect. \ref{iresect}, the energy released by accreting some matter, can unbind additional matter via radiation pressure. The highest spherical accretion rate possible without the arriving matter stopped and blown outwards by radiation pressure is the Eddington accretion limit $\dot{M}_\mathrm{Edd} = 4 \pi c R_\mathrm{a}/\kappa_\mathrm{es}$, with $c$ the speed of light, $R_\mathrm{a}$ the accretor radius and $\kappa_\mathrm{es} \approx 0.199\ \mbox{cm}^{2} \mbox{g}^{-1}$ the electron Thomson scattering mean opacity for hydrogen-deficient matter. The maximum mass transfer rate that can be survived via isotropic re-emission is given by the ratio of the potential well at the accretor surface $\Phi_\mathrm{a}$ and the potential in the first Lagrangian point \citep{begelman1979,king1999,yungelson2002}. The potential in L1 by definition is equal to the corotating-binary potential at the Roche lobe surface $\Phi_\mathrm{L}$, which can be approximated to within $0.15\%$ by

\begin{equation} 
    \label{pot}
    \frac{- \Phi_\mathrm{L}}{G M_\mathrm{tot}/a} = \frac{3}{2} + \exp{\left\{ \frac{\ln(10)}{3} - \left( \frac{12}{5} + \left( \frac{4}{9} \ln^{2}(q) \right)^{9/8} \right)^{7/16} \right\}}
\end{equation}
with $0 < q < \infty$, $M_\mathrm{tot} = M_\mathrm{a} + M_\mathrm{d}$ the total binary mass and $G$ the gravitational constant.
If we define $\phi(q)$ as the right-hand side of Eq. (\ref{pot}), the maximum survivable mass transfer rate is

\begin{equation}
    \label{eqire}
    \dot{M}_\mathrm{mir} = \dot{M}_\mathrm{Edd} \cdot \frac{\Phi_\mathrm{a}}{\Phi_\mathrm{L}} = \frac{4 \pi c R_\mathrm{a}}{\kappa_\mathrm{es}} \cdot \frac{-G M_\mathrm{a} / R_\mathrm{a}}{- \phi(q) G M_\mathrm{tot} / a} = \frac{4 \pi c}{\kappa_\mathrm{es}} \cdot \frac{a}{\phi(q)(1+q)}
\end{equation}
and is independent of $R_\mathrm{a}$.
Since the matter is ejected from the vicinity of the accretor \citep{soberman1997,tauris1999}, it has approximately the same specific orbital angular momentum as the accretor.

\subsection{Response of Roche-lobe radius to mass transfer}
\label{rlrad}

Any effect of change in angular momentum because of mass transfer and ejection from the system is contained in $\zeta_\mathrm{L} = \partial \ln R_\mathrm{L}/\partial \ln M_\mathrm{d}$ (Sect. \ref{appmt}).
The effective radius of the donor Roche lobe $R_\mathrm{L}$ can be approximated to within $1\%$ by \citet{eggleton1983}

\begin{equation}
    \label{roche}
    \frac{R_\mathrm{L}}{a} = \frac{0.49}{0.6 + q^{-2/3}\ln(1+q^{1/3})}.
\end{equation}
Given $q = M_\mathrm{d}/M_\mathrm{a}$ and $R_\mathrm{d}$, Eqs. (\ref{basic}) (first part) and (\ref{roche}) yield $a$. The Roche-lobe geometry only applies when the spin period of the Roche-lobe filling component is equal to the orbital period, and when the orbit is circular. These conditions are normally met in compact Roche-lobe filling systems.

Logarithmic differentiation of $R_\mathrm{L}$ yields
\begin{equation}
    \label{zetarl}
    \zeta_\mathrm{L} = \zeta_{a} + \zeta_{r_\mathrm{L}},
\end{equation}
where $\zeta_{a} \equiv \partial \ln a/\partial \ln M_\mathrm{d}$ and

\begin{eqnarray}
    \label{zetae}
    \zeta_{r_\mathrm{L}} &\equiv& \frac{\partial \ln (R_\mathrm{L}/a)}{\partial \ln M_\mathrm{d}} \nonumber \\
                     &=& \zeta_{q} \cdot \frac{1}{3} \left[ \frac{2\ln(1+q^{1/3}) - (1+q^{-1/3})^{-1}}{\ln(1+q^{1/3}) + 0.6q^{2/3}} \right]
\end{eqnarray}
with

\begin{equation}
    \label{zetaq2}
    \zeta_{q} \equiv \partial \ln q / \partial \ln M_\mathrm{d} = 1 + (1-\beta)q,
\end{equation}
where $\beta$ is the fraction of mass lost by the donor that leaves the system via isotropic re-emission, see Eq. (\ref{beta}). $\zeta_{r_\mathrm{L}}$ with $\beta = 0$ is the same as in \citet{marsh2004}, who studied mass transfer in binary white dwarfs.

\subsubsection{Mass flows}
\label{flows}

In order to determine $\zeta_{a}$, we consider three mass flows in the binary system: mass lost by the donor can either be accreted, stay in the disk or be unbound from the system.

\paragraph{Accreted mass}

\begin{figure}
\resizebox{\hsize}{!}{\includegraphics{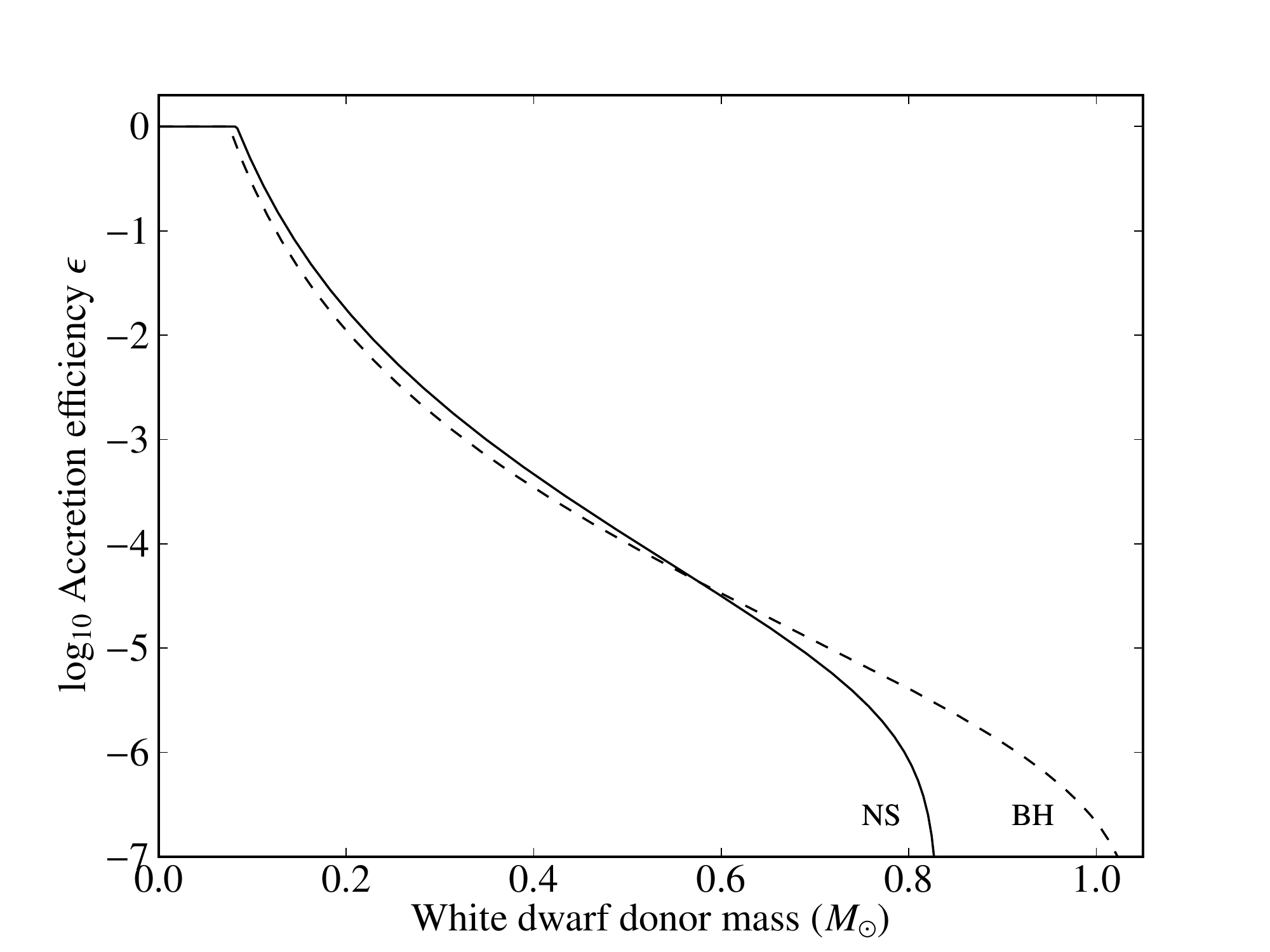}} 
\caption{Accretion efficiency $\epsilon$ (Eq. \ref{epsilon}) for a white dwarf donor and an initially $1.4\ M_{\odot}$ neutron star (solid) and a $10\ M_{\odot}$ black hole (dashed) accretor, in the case of full feedback of angular momentum. Both curves start at their respective dynamical stability limit (Sect. \ref{dscrit}).}
\label{fig:epsilon}
\end{figure}

The accretion efficiency $\epsilon$ (where $0 \le \epsilon \le 1$) is the fraction of the matter lost by the donor that is accreted by the primary

\begin{equation}
    \label{noncons}
    \dot{M}_\mathrm{a} = -\epsilon \dot{M}_\mathrm{d}.
\end{equation}
When we assume full accretion in the case of a sub-Eddington mass transfer rate (in the absence of magnetic fields or instabilities)

\begin{equation}
    \label{epsilon}
    \epsilon = \min \, (-\frac{\dot{M}_\mathrm{Edd}}{\dot{M}_\mathrm{d}},1).
\end{equation}
$\dot{M}_\mathrm{Edd} \approx 3 \cdot 10^{-8}\ M_{\odot}\mbox{yr}^{-1}$ for a neutron star that is accreting hydrogen-deficient matter. The evolution of $\epsilon$ for two different accretor masses is illustrated in Fig. \ref{fig:epsilon}.

\paragraph{Mass lost from the system}

In the case of non-conservative mass transfer the system loses angular momentum along with the ejected matter. The specific angular momenta of the ejected matter and the binary are $\dot{J}_\mathrm{orb}/\dot{M}_\mathrm{tot}$ and $J_\mathrm{orb}/M_\mathrm{tot}$, respectively. The parameter $\gamma$ is defined as the ratio between these

\begin{equation}
    \gamma = \frac{(\dot{J}/J)_\mathrm{orb}}{\dot{M}_\mathrm{tot}/M_\mathrm{tot}},
\end{equation}
(note that $\gamma$ in \citet{soberman1997} has a different meaning, unlike $\beta$ and $\epsilon$). It can easily be derived that $\gamma = q$ in the case of isotropic re-emission, because matter has the specific orbital angular momentum of the accretor (Sect. \ref{ire}).

\paragraph{Mass added to accretion disk}

In the absence of an accretion disk that can efficiently transfer angular momentum back to the orbit, the stream and torque terms in Eq. (\ref{j_balance}) do not cancel out, and the disk becomes a net sink of angular momentum. In the extreme case of no feedback at all $\dot{J}_\mathrm{torque} = 0$, leaving $\dot{J}_\mathrm{stream} = \sqrt{G M_\mathrm{a} R_\mathrm{h}} \dot{M}_\mathrm{d}$ \citep{hut1984}. The circularization radius $R_\mathrm{h}$ is the orbital distance at which the specific angular momentum is the same as in the incoming stream from the donor \citep{osaki1989}, minus the amount of angular momentum removed by the donor, approximated by \citep{verbunt1988}

\begin{eqnarray}
    \label{verbuntfit}
    \frac{R_\mathrm{h}}{a} &=& 0.0883 - 0.04858 \log_{10}(q) \nonumber \\
        && + 0.11489 \log_{10}^{2}(q) + 0.020475 \log_{10}^{3}(q),
\end{eqnarray}
adjusted for our definition of $q = M_\mathrm{d}/M_\mathrm{a}$, and valid for $10^{-3} < q < 1$.

In the general case of limited feedback, we introduce a factor $\eta$ (where $0 \le \eta \le 1$) describing the fraction of mass lost from the donor that is (net) added to the disk because it cannot be accreted, defined analogous to $\epsilon$ in Eq. (\ref{noncons})

\begin{equation}
    \dot{M}_\mathrm{disk} = -\eta \dot{M}_\mathrm{d}.
\end{equation}

Conservation of mass is represented by $\beta + \epsilon + \eta = 1$. In the case of full feedback ($\eta = 0$), $\beta$ may be estimated by assuming that all non-accreted matter leaves the system, unless the maximum isotropic re-emission rate $\dot{M}_\mathrm{mir} > 0$ (Eq. \ref{eqire}) is too low to accomplish this, so

 \begin{equation}
    \label{beta}
    \beta = \min \, (-\frac{\dot{M}_\mathrm{mir}}{\dot{M}_\mathrm{d}},1-\epsilon).
\end{equation}

\subsubsection{Response of semi-major axis to mass transfer}

Combining the above equations, the change in angular momentum of the system $\dot{J}_\mathrm{orb}$ because of mass leaving the system or entering the disk is

\begin{eqnarray}
    \label{specam2}
    \left( \frac{\dot{J}}{J} \right)_\mathrm{orb} &=& \gamma \frac{\dot{M}_\mathrm{tot}}{M_\mathrm{tot}} + \eta \frac{\sqrt{G M_\mathrm{a} R_\mathrm{h}} \dot{M}_\mathrm{d}}{J_\mathrm{orb}} \nonumber \\
        &=& \gamma \frac{\dot{M}_\mathrm{tot}}{M_\mathrm{tot}} + \eta \sqrt{(1+q)r_\mathrm{h}} \frac{\dot{M}_\mathrm{d}}{M_\mathrm{d}},
\end{eqnarray}
with $r_\mathrm{h} = R_\mathrm{h}/a$.

To find $\zeta_{a}$, we first differentiate the logarithm of the equation for orbital angular momentum to time. We introduce a disk mass term to allow for transferred matter to stay in the disk instead of being accreted, corresponding to possible limited feedback to find

\begin{equation}
    \label{orbangdiff2}
    \left( \frac{\dot{J}}{J} \right)_\mathrm{orb} = \frac{1}{2}\frac{\dot{a}}{a} + \frac{\dot{M}_\mathrm{d}}{M_\mathrm{d}} + \frac{\dot{M}_\mathrm{a}+\dot{M}_\mathrm{disk}}{M_\mathrm{a}} - \frac{1}{2}\frac{\dot{M}_\mathrm{tot}}{M_\mathrm{tot}},
\end{equation}
where $M_\mathrm{disk} \ll M_\mathrm{a}$ has been used.
Mass added to a disk around the accretor has the same effect on $J_\mathrm{orb}$ as mass that is accreted, because the accretor-disk subsystem is orbiting the center of mass of the binary. Hence the generalization of accretor mass to include disk mass.

Equating Eqs. (\ref{specam2}) and (\ref{orbangdiff2}) eliminates $(\dot{J}/J)_\mathrm{orb}$. Inserting Eq. (\ref{noncons}), $\dot{M}_\mathrm{tot} = \beta \dot{M}_\mathrm{d}$ and $\dot{M}_\mathrm{a}+\dot{M}_\mathrm{disk} = (\beta-1)\dot{M}_\mathrm{d}$ gives

\begin{equation}
    \label{zetaa2}
    \zeta_{a} = -2 \cdot \left[ 1 - (1-\beta) q - \frac{\beta (\gamma+\frac{1}{2})}{1 + q^{-1}} - \eta \sqrt{(1+q)r_\mathrm{h}} \right].
\end{equation}

In the case of full feedback, $\eta = 0$ and $\beta + \epsilon = 1$.
At the onset of mass transfer from a massive donor, when a disk has not yet formed, $\eta = 1$ and $\zeta_{a} = -2[1-q-\sqrt{(1+q)r_\mathrm{h}}]$, identical to the equation in \citet{verbunt1988,marsh2004}. In the case of low-mass donors where accretion is made impossible by limited feedback, accretion has ceased ($\epsilon = 0$) and isotropic re-emission has stopped accordingly ($\beta = 0$), hence $\eta = 1$.
The actual value of $\eta$ during the evolution of an UCXB will be considered in Sect. \ref{reso}.

The $\zeta_\mathrm{L}$ relation as a function of $q$ for $\epsilon = 0$ is shown in Fig. \ref{fig:zetas_simple} for two accretor masses and for more general mass flow parameters in Fig. \ref{fig:zetas}, for only a neutron star accretor. The Roche-lobe radius depends on both component masses and angular momentum losses via its dependence on $a$, and on the mass ratio via $R_\mathrm{L}/a$.

Note that the variables $\epsilon$, $\dot{M}_\mathrm{d}$, $\zeta_\mathrm{L}$, $(\dot{J}/J)_\mathrm{\textsc{gwr}}$, $M_\mathrm{d}$, $M_\mathrm{a}$ and $\dot{M}_\mathrm{Edd}$, and in the case of a black hole accretor also $R_\mathrm{a}$, are circularly related to each other through Eq. (\ref{mt2}), the gravitational wave equation, and Eqs. (\ref{epsilon}) and (\ref{zetarl}). Hence, iteration at each mass step is necessary to arrive at the equilibrium values.

\subsection{White dwarf radius}
\label{wdrad}

Now $\zeta_\mathrm{L}$ is known, only the response of the white dwarf donor radius to mass loss, $\zeta_\mathrm{d}$, is still needed to evaluate Eq. (\ref{mt2}).
We justify the use of a zero-temperature white dwarf mass-radius relation by noting that thermal pressure is negligible for significantly degenerate objects such as white dwarfs with a mass above $0.01\ M_{\odot}$, and still quite small for lower masses.
From the mass-radius relations derived by \citet{deloye2003} it follows that the effect of ideal gas pressure on the radius, additional to degeneracy pressure and Coulomb attraction, is less than $10\%$ for $M_\mathrm{d} > 10^{-2}\ M_{\odot}$ when the temperature $T = 10^{6}\ \mbox{K}$, less than $2.3\%$ for $M_\mathrm{d} > 4 \cdot 10^{-3}\ M_{\odot}$ when $T = 10^{5}\ \mbox{K}$ and less than $0.1\%$ for $M_\mathrm{d} > 10^{-3}\ M_{\odot}$ when $T = 10^{4}\ \mbox{K}$. Temperature is important only for very low-mass or very hot donors. For a mass ratio below $\sim 0.01$, donor radii could be $10-30\%$ higher than for the zero-temperature case \citep{nelson2003}.
Furthermore, we consider only degenerate donors, since any fusion will be extinguished on a relatively short timescale after some mass has been lost \citep{savonije1986,yungelson2008}. Also, a non-degenerate outer layer can have a significant effect on the radius response to mass loss, but again this is temporary -- once the layer has been lost, a degenerate object remains.

The radius $R_\mathrm{d}$ of a zero-temperature white dwarf (the ratio of atomic number to atomic weight $Z/A = 1/2$) of mass $M_\mathrm{d}$ by Eggleton \citep{rappaport1987} is

\begin{eqnarray}
    \label{rwd}
    \frac{R_\mathrm{d}}{R_{\odot}}   &=& 0.0114 \left[ \left(\frac{M_\mathrm{d}}{M_\mathrm{Ch}}\right)^{-2/3} - \left(\frac{M_\mathrm{d}}{M_\mathrm{Ch}}\right)^{2/3} \right]^{1/2} \nonumber \\
                    && \cdot \left[ 1 + \frac{7}{2} \left(\frac{M_\mathrm{d}}{M_\mathrm{p}}\right)^{-2/3} + \left(\frac{M_\mathrm{d}}{M_\mathrm{p}}\right)^{-1} \right]^{-2/3}
\end{eqnarray}
with $M_\mathrm{Ch} = 1.44\ M_{\odot}$. For a helium white dwarf, $M_\mathrm{p} = 5.66 \cdot 10^{-4}\ M_{\odot}$ and for a carbon-oxygen white dwarf (assuming 50\% of each element) $M_\mathrm{p} = 1.44 \cdot 10^{-3}\ M_{\odot}$, valid for $0 < M_\mathrm{d} < M_\mathrm{Ch}$ \citep{marsh2004}. $\zeta_\mathrm{d} \equiv \partial \ln R_\mathrm{d}/\partial \ln M_\mathrm{d}$ is determined by logarithmic differentiation

\begin{equation}
    \label{zetawd}
    \zeta_\mathrm{d} = -\frac{1}{3} \left[ \frac{1 + \left(\frac{M_\mathrm{d}}{M_\mathrm{Ch}}\right)^{4/3}}{1 - \left(\frac{M_\mathrm{d}}{M_\mathrm{Ch}}\right)^{4/3}} \right] + \frac{2}{3} \left[ \frac{ 1+ \frac{7}{3} \left(\frac{M_\mathrm{d}}{M_\mathrm{p}}\right)^{1/3}}{1 + \frac{7}{2}\left(\frac{M_\mathrm{d}}{M_\mathrm{p}}\right)^{1/3} + \left(\frac{M_\mathrm{d}}{M_\mathrm{p}}\right)} \right]
\end{equation}
and is illustrated by the dashed curve in Fig. \ref{fig:zetas_simple}. Throughout this paper we consider helium white dwarfs, though all results except quantitative details and the disk instability details also apply to carbon-oxygen composition.

\subsection{Propeller effect}
\label{prop}

\subsubsection{Magnetic field of neutron star accretors}

Neutron stars typically possess a strong magnetic field. During accretion, the magnetic field strength is reduced \citep{bhattacharya1991}, but even for old millisecond pulsars with a spin period below $30\ \mbox{ms}$, it can still be $10^{8-9}\ \mbox{G}$ \citep{wang2011}. The magnetic field which is locked to the neutron star is supposed to penetrate the accretion disk, and in the case of a rapidly spinning neutron star, the outer magnetosphere forces the orbiting matter to velocities exceeding the local Kepler velocity by exerting a torque. As a result of the angular momentum gained, disk material in this region moves outwards, which reduces accretion and can lead to unbinding the matter from the system.

\subsubsection{Critical mass transfer rate}
\label{critmt}

The Alfv\'{e}n radius (also known as magnetosphere radius) is the distance from the neutron star below which the magnetic energy density (i.e. magnetic pressure) dominates the ram pressure in the accretion disk \citep{lamb1973}.

\begin{equation}
    \label{energy}
    \frac{B^{2}}{8\pi} = \rho v_\mathrm{K}^{2},
\end{equation}
with $B = B_\mathrm{NS} \cdot (R_\mathrm{NS}/R)^{3}$ the magnetic induction of the neutron star (NS) in vacuum, $B_\mathrm{NS}$ the stellar surface field at the magnetic equator, $\rho$ the average density at radius $R$ and $v_\mathrm{K} = \sqrt{GM_\mathrm{NS}/R}$ the Keplerian orbital velocity.

We use the spherical continuity equation (see justification in Sect. \ref{diskacc})

\begin{equation}
    \label{conteq}
    \dot{M}_\mathrm{NS} = 4\pi R^{2} \rho v_\mathrm{ff},
\end{equation}
with $v_\mathrm{ff} = \sqrt{2} v_\mathrm{K}$ the free-fall velocity, to eliminate $\rho$ and solve for the Alfv\'{e}n radius \citep{davidson1973,elsner1977}

\begin{equation}
    \label{alfven}
    R_{\mu} = \left( \frac{\mu^{4}}{2GM_\mathrm{NS}\dot{M}_\mathrm{NS}^{2}} \right)^{1/7},
\end{equation}
where $\mu = B_\mathrm{NS} R_\mathrm{NS}^{3}$ is the magnetic dipole moment of the neutron star.

Since the magnetic field lines are dragged along with the neutron star's rotation, their orbital angular frequency equals the spin angular frequency $\omega_\mathrm{s}$ of the neutron star. At a distance $R$ from the neutron star, the corotating field lines have an orbital velocity $\omega_\mathrm{s} R$. This equals the Keplerian velocity at the corotation radius

\begin{equation}
    \label{corot}
    R_\mathrm{co} = \left( \frac{GM_\mathrm{NS}}{\omega_\mathrm{s}^{2}} \right)^{1/3}.
\end{equation}
The propeller effect acts only if the magnetic field energy dominates the kinetic energy of the disk beyond the corotation radius, and when the orbital velocity of the field lines exceeds the escape velocity at the Alfv\'en radius \citep{rappaport2004}, i.e. $\omega_\mathrm{s} R_{\mu} > \sqrt{2} v_\mathrm{K,\mu}$, where $v_\mathrm{K,\mu}$ is the Kepler velocity at $R_{\mu}$, so $\omega_\mathrm{s} > \sqrt{2} \omega_\mathrm{K,\mu}$ which implies $R_{\mu} > 2^{1/3} R_\mathrm{co}$. Inserting Eqs. (\ref{alfven}) and (\ref{corot}) yields the propeller criterion

\begin{equation}
    \label{mdotprop}
    \dot{M}_\mathrm{NS} < \frac{\mu^{2} \omega_\mathrm{s}^{7/3}}{(2GM_\mathrm{NS})^{5/3}}.
\end{equation}

\subsubsection{Disk accretion}
\label{diskacc}

By using the thin disk description by \citet{dunkel2006} it follows that the magnetic field is far too weak to dominate the gas flow in the disk even for the lowest expected accretion rates. Even for the slightly unrealistic case of a neutron star with a $10^{9}$ G equatorial magnetic field and a $1$ ms spin period, the magnetic field only dominates the kinetic energy of the Kepler flow at the corotation radius if the mass transfer rate is lower than $10^{-16}\ M_{\odot} \mbox{yr}^{-1}$,
$\sim 10^{10}$ times lower than for the spherically infalling case.
High density blobs of gas can also dominate the magnetic field and be accreted \citep{aly1990}. Such blobs could form if matter piles up near the Alfv\'en radius, held back by the magnetic field, and enter once enough matter has accumulated.

The conclusion is that Eq. (\ref{alfven}) is an upper limit for the Alfv\'en radius since the accretion is not spherical as assumed in Eq. (\ref{conteq}) but occurs in a
disk. However, since the density and the geometry of the interacting disk-magnetosphere region is poorly known, we choose to allow for disk accretion in the equations by also considering an effective magnetic field strength at the stellar surface which is smaller than the real field strength $B_\mathrm{NS}$.

\subsection{Disk Instability Model}
\label{dim}

\citet{zand2007} make a rough estimate for the lowest stable mass transfer rate for helium composition based on \citet{lasota2001} and \citet{menou2002}

\begin{equation}
    \label{mdotdim}
    \dot{M}_\mathrm{crit} \approx 3 \cdot 10^{-10} \left( \frac{M_\mathrm{a}}{M_{\odot}} \right)^{0.3} \left( \frac{P_\mathrm{orb}}{\mbox{hr}} \right)^{1.4} M_{\odot} \mbox{yr}^{-1}
\end{equation}
where $P_\mathrm{orb}$ is the orbital period. Once the mass transfer rate decreases below this value, a thermal-viscous instability appears (Sect. \ref{dim_intro}) and the disk experiences outbursts.
The timescale on which an accretion disk rebuilds after a collapse is equal to $M_\mathrm{disk}/\dot{M}_\mathrm{d}$, where the disk mass is given by

\begin{equation}
    \label{diskmass}
    M_\mathrm{disk} = \int_{R_\mathrm{in}}^{R_\mathrm{out}} 2\pi R \Sigma \mbox{d}R,
\end{equation}
with $\Sigma$ the surface density for zone C in \citet{dunkel2006}. $R_\mathrm{in}$ and $R_\mathrm{out}$ are the inner and outer disk radii and will be estimated later.

The Disk Instability Model is still not fully understood, as illustrated by the case of \object{SS Cygni}, a dwarf nova that according to the Disk Instability Model is too bright and hot to have an unstable disk \citep{schreiber2007}. This increases the uncertainty in the mass transfer rate given by Eq. (\ref{mdotdim}). On the other hand, the intermediate state in AM CVn systems seems to be predicted quite well \citep{tsugawa1997}.

\section{Results}
\label{results}

\subsection{Feedback of angular momentum}
\label{reso}

Since $\eta$ is the most uncertain mass flow parameter in Sect. \ref{flows}, here we first look into the likelihood of limited feedback of angular momentum at some low mass ratio. If the torque between outer disk and donor is too weak, a dynamical instability may result (Sect. \ref{explain_feedback}).

\begin{figure}
\resizebox{\hsize}{!}{\includegraphics{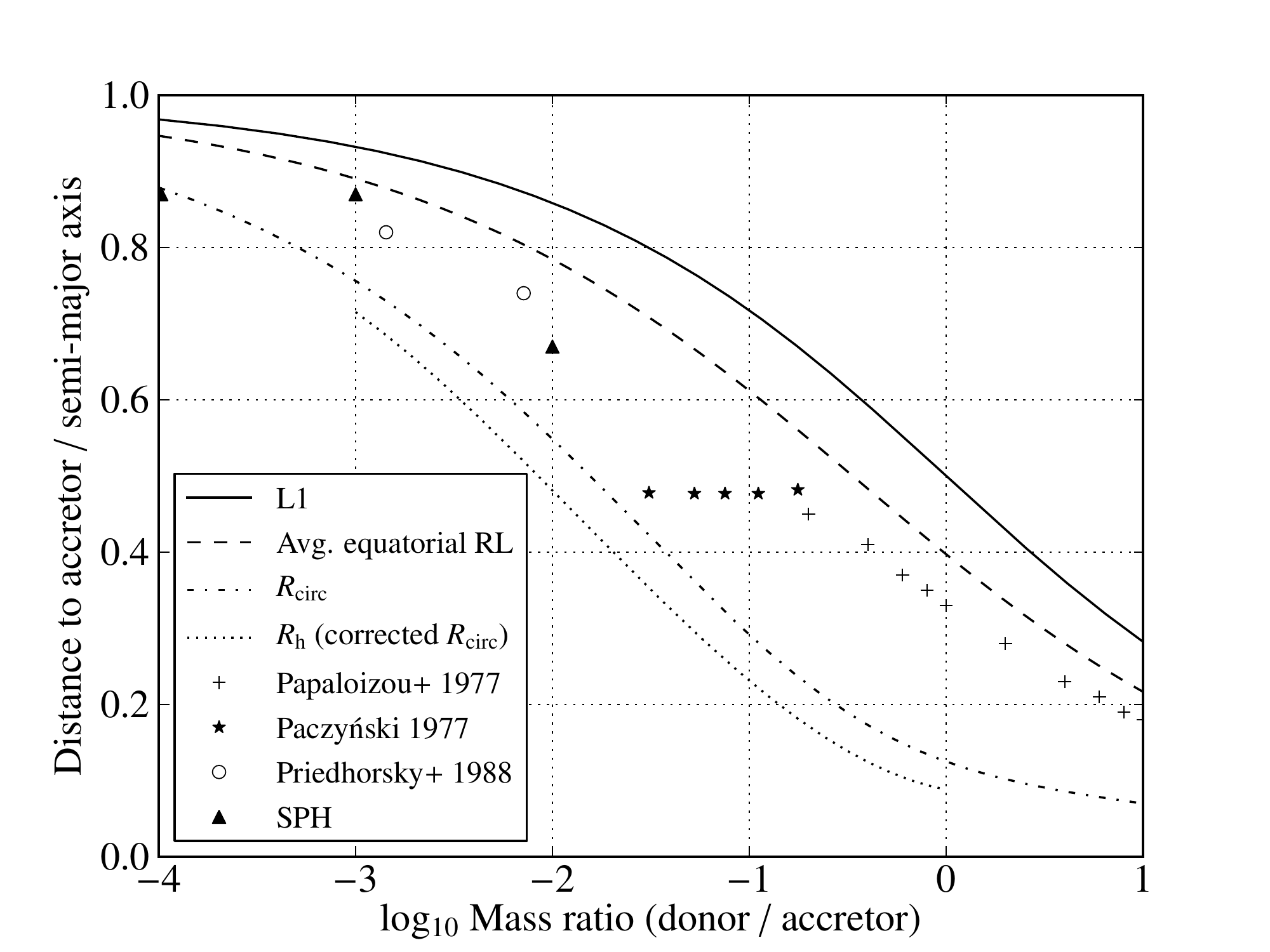}} 
\caption{Important radii within the accretion disk; the distance between the accretor and the L1 point $d_\mathrm{L1}$ (solid), the average equatorial Roche lobe radius (i.e. in the orbital plane) (dashed), the circularization radius $R_\mathrm{circ}$ (dash-dotted) and $R_\mathrm{circ}$ corrected for donor influence during free-fall, $R_\mathrm{h}$ (dotted, Eq. \ref{verbuntfit}). Furthermore, the pluses (+) indicate the outer tidally stable radius from \citet{papaloizou1977}. The stars ($\star$) indicate the largest orbit that does not intersect others, according to \citet{paczynski1977} (which lies near the 3:1 resonance). The two open circles are outer disk estimates by \citet{priedhorsky1988}, and the three filled triangles are outer disk radii from our SPH simulations.}
\label{fig:radii}
\end{figure}

Also mentioned in Sect. \ref{explain_feedback} was that \citet{yungelson2006} and \citet{lasota2008} have suggested that a resonance between the orbital period and the (mean) period of particles in the disk may prevent the accretion disk from redistributing angular momentum. Also, when $q \lesssim 0.02$, the circularization radius exceeds the estimates of the outer radius by \citet{paczynski1977} and \citet{papaloizou1977} ($\sim 0.48a$, see the stars and pluses in Fig. \ref{fig:radii}) which could hinder accretion, and at the same time reduce the feedback of angular momentum, making the binary less stable. Their value of $q = 0.02$ can be refined by using the adjusted circularization radius equation by \citet{verbunt1988} (dotted curve in Fig. \ref{fig:radii}), who correct the circularization radius $R_\mathrm{circ}/a = (1+q)(d_\mathrm{L1}/a)^{4}$ \citep{frank2002book} (dash-dotted) for the effect of the donor on the stream of matter. During this process, angular momentum is extracted from the stream, so $R_\mathrm{h} < R_\mathrm{circ}$. Using $R_\mathrm{h}$ (Eq. \ref{verbuntfit}) instead of $R_\mathrm{circ}$ lowers the critical mass ratio to $q = 0.01$ (Fig. \ref{fig:radii}).
Figure \ref{fig:radii} also shows that the location of the first Lagrangian point (solid curve) is very close to the donor at low mass ratio, and that the equatorial Roche lobe radius (dashed) also approaches the L1 point, because the equatorial accretor Roche lobe becomes more circular rather than teardrop-shaped at low mass ratio.\footnote{Note that the equatorial Roche lobe radius is significantly larger than the average radius over the whole volume (Eq. \ref{roche}), similar to an equatorial bulge.}

\begin{figure}
\resizebox{\hsize}{!}{\includegraphics{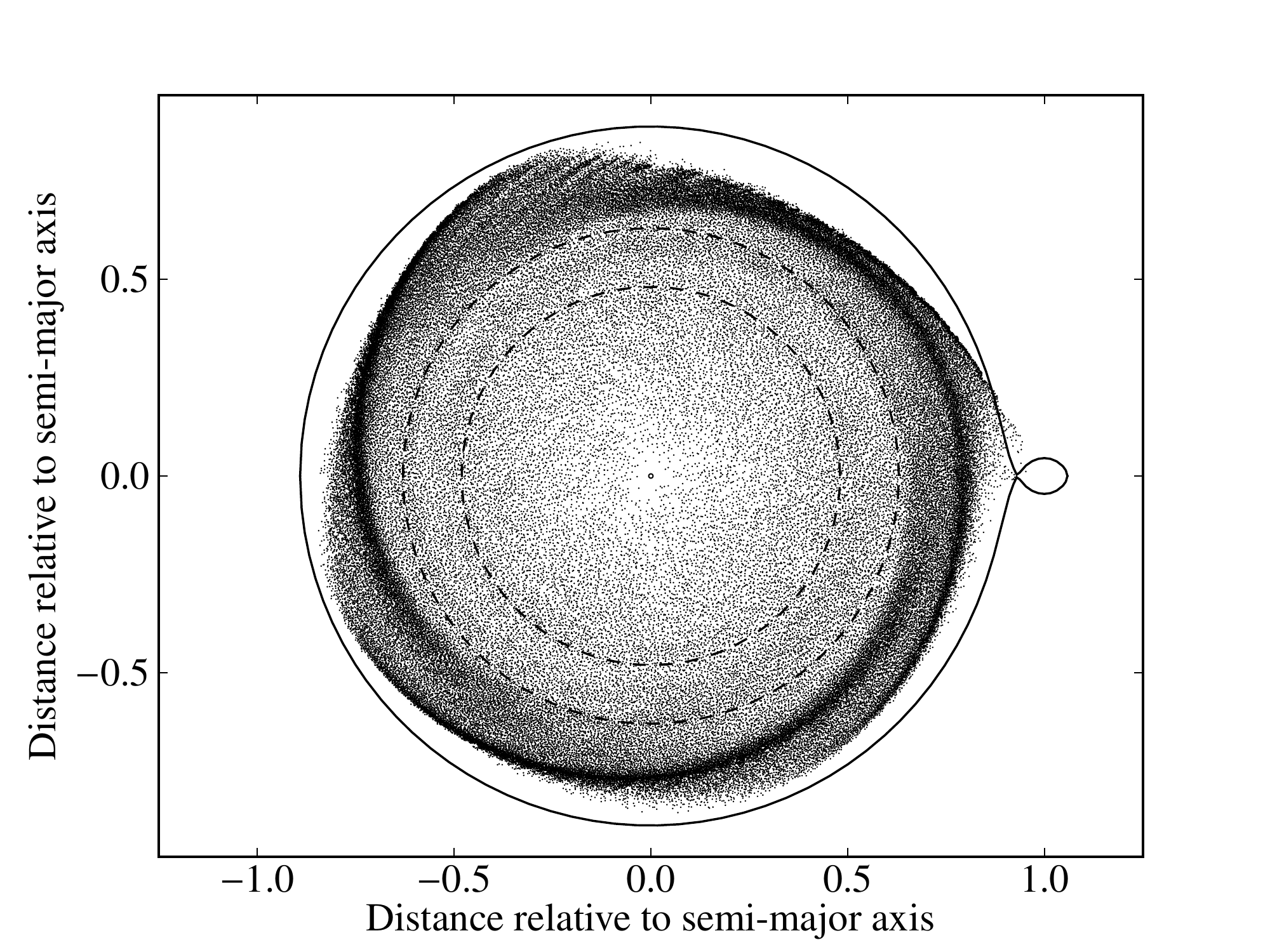}}
\caption{Numerical smoothed particle hydrodynamics simulation of an accretion disk in dynamical equilibrium for mass ratio $10^{-3}$ after 400 orbits. The solid curve represents the equatorial Roche lobes. The accretor has been marked by a small open circle. The dashed circles indicate the 3:1 (inner) and 2:1 (outer) resonances with the orbital period.}
\label{fig:matt}
\end{figure}

We ran numerical smoothed particle hydrodynamics (SPH) simulations such as that in Fig. \ref{fig:matt} which show that disks function properly even at a very low mass ratio. The figure shows a disk in dynamical equilibrium after 400 orbits (see \citet{simpson1998} and \citet*{wood2009} for details of the numerics). The simulations show that the disk fits well inside the Roche lobe (triangles in Fig. \ref{fig:radii}). This confirms the conclusion of \citet{priedhorsky1988}, who used an accretion disk model to find that the disk expansion is not stopped at $0.48a$ and that the disk stays within the accretor Roche lobe (circles in Fig. \ref{fig:radii}) and tidal torques are strong enough to transfer back all angular momentum.
Also, nothing special appears to happen in the disk near the most important resonant radii, the circles in Fig. \ref{fig:matt}. Hence, $\eta$ remains zero at very low mass ratio. In the remainder of the paper we will consider the full-feedback case, except for Sect. \ref{res_inst}, where we will show evolutionary tracks for the case in which feedback somehow stops at a low mass ratio.

Observations of binaries cannot rule out the existence of a dynamical instability since no systems with an upper mass ratio below $0.01$ have been discovered, although a couple have minimum mass ratios below this limit \citep{galloway2002,krimm2007,altamirano2010}.

\subsection{The evolution of ultracompact X-ray binaries}
\label{tracks}

The key elements in understanding the evolution of UCXBs are represented by Eq. (\ref{mt2}). Mass transfer is driven by gravitational wave radiation, and also depends on the relative responses of donor and Roche lobe size to mass transfer. In Sect. \ref{reso} we found that advected angular momentum will be returned from disk to orbit for any mass ratio, which constrains $\zeta_\mathrm{L}$ and allows us to solve all relevant binary parameters during the entire evolution when assuming stable accretion, rather than in outbursts.
A system can survive the onset of mass transfer only if the accretor can eject all transferred matter exceeding the Eddington limit via isotropic re-emission. For any realistic stellar-mass accretor, a merger can be avoided if the white dwarf donor has a mass below $\sim 0.3\ M_{\odot}$.
Mass transfer is dynamically stable if, upon mass loss, the donor expands less than its Roche lobe (Eq. \ref{stab}), that is, when the solid curve in Fig. \ref{fig:zetas} lies below the dashed curve.
UCXBs with black hole accretors can theoretically exist, but may be rarer than neutron star UCXBs due to the steep slope of the initial mass function. However, the binary survival rate during the supernova event can be different for black hole formation than it is for neutron star formation due to a different kick velocity distribution.

\subsubsection{Evolution in the case of full feedback}
\label{acc}

\begin{figure}
\resizebox{\hsize}{!}{\includegraphics{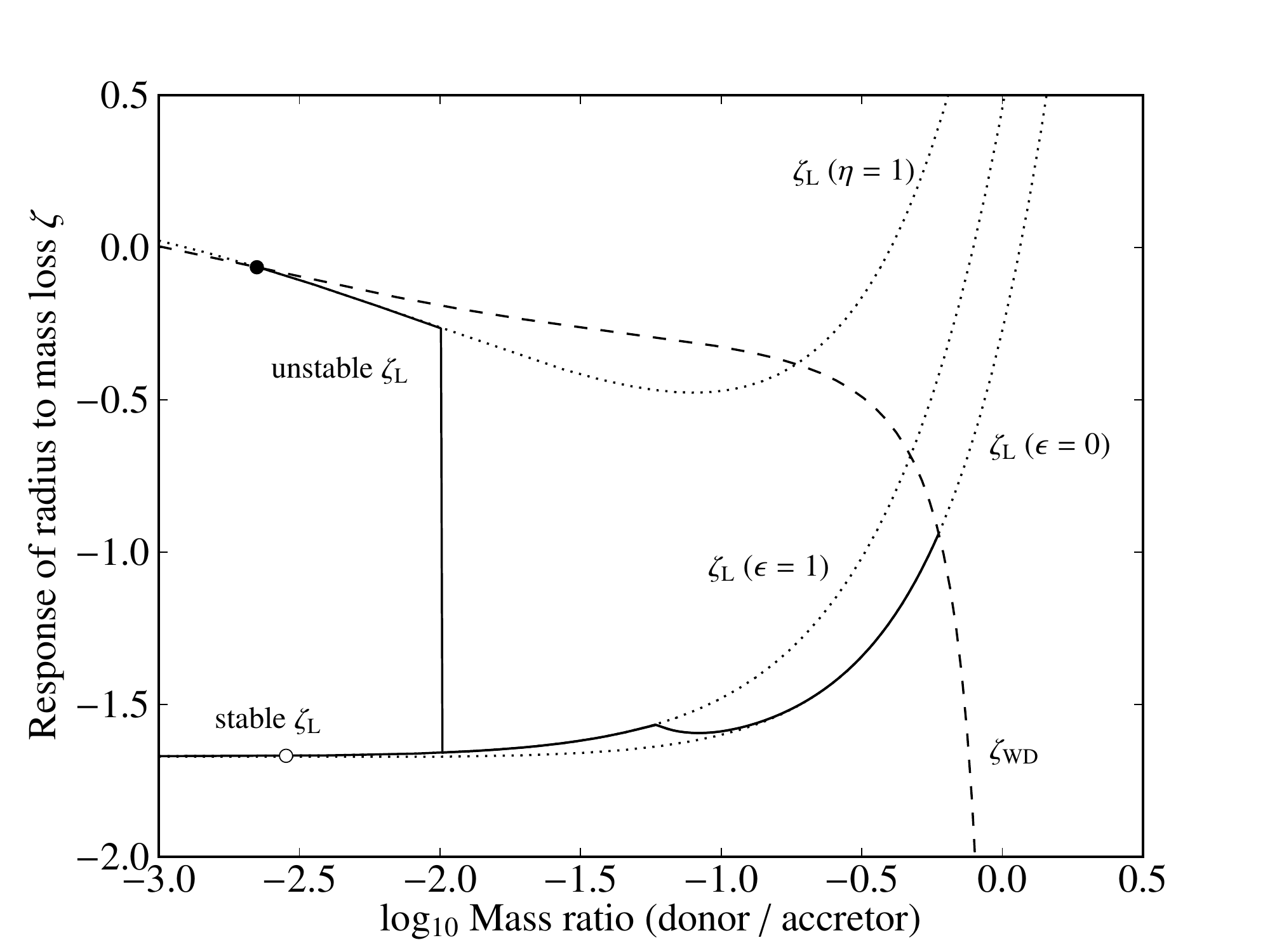}} 
\caption{Mass-radius relation exponent $\zeta_\mathrm{L}$ (Sect. \ref{rlrad}) for various modes of mass transfer from a white dwarf to a neutron star of initially $1.4\ M_{\odot}$ with a $12\ \mbox{km}$ radius. The lower dotted curve shows the case of accretion efficiency $\epsilon = 0$ and full feedback ($\eta = 0$) (almost all transferred matter lost via isotropic re-emission, hardly any accretion). The middle dotted curve is for $\epsilon = 1$ and $\eta = 0$ (conservative mass transfer). The upper dotted curve is for no feedback ($\eta = 1$). The zero-temperature white dwarf exponent $\zeta_\mathrm{d}$ (Sect. \ref{wdrad}) is represented by the dashed curve. The solid curves starting at $q = 0.59$ ($M_\mathrm{d} = 0.83\ M_{\odot}$, Sect. \ref{dscrit}) indicate the actual value of $\zeta_\mathrm{L}$ (for $\epsilon$ shown in Fig. \ref{fig:epsilon}) for both the full feedback and no-feedback case. Until $q = 0.01$ there is full feedback in either case. Below $q = 0.01$ the full feedback case is shown by the lower solid branch, where the open disk denotes an age of $10$ Gyr (since the onset of mass transfer). The no-feedback case has a jump in $\zeta_\mathrm{L}$ at $q = 0.01$, after which it follows the upper dotted curve until the intersection with the dashed curve (filled disk) where donor disruption occurs at a system age of $300$ Myr.}
\label{fig:zetas}
\end{figure}

\begin{figure}
\resizebox{\hsize}{!}{\includegraphics{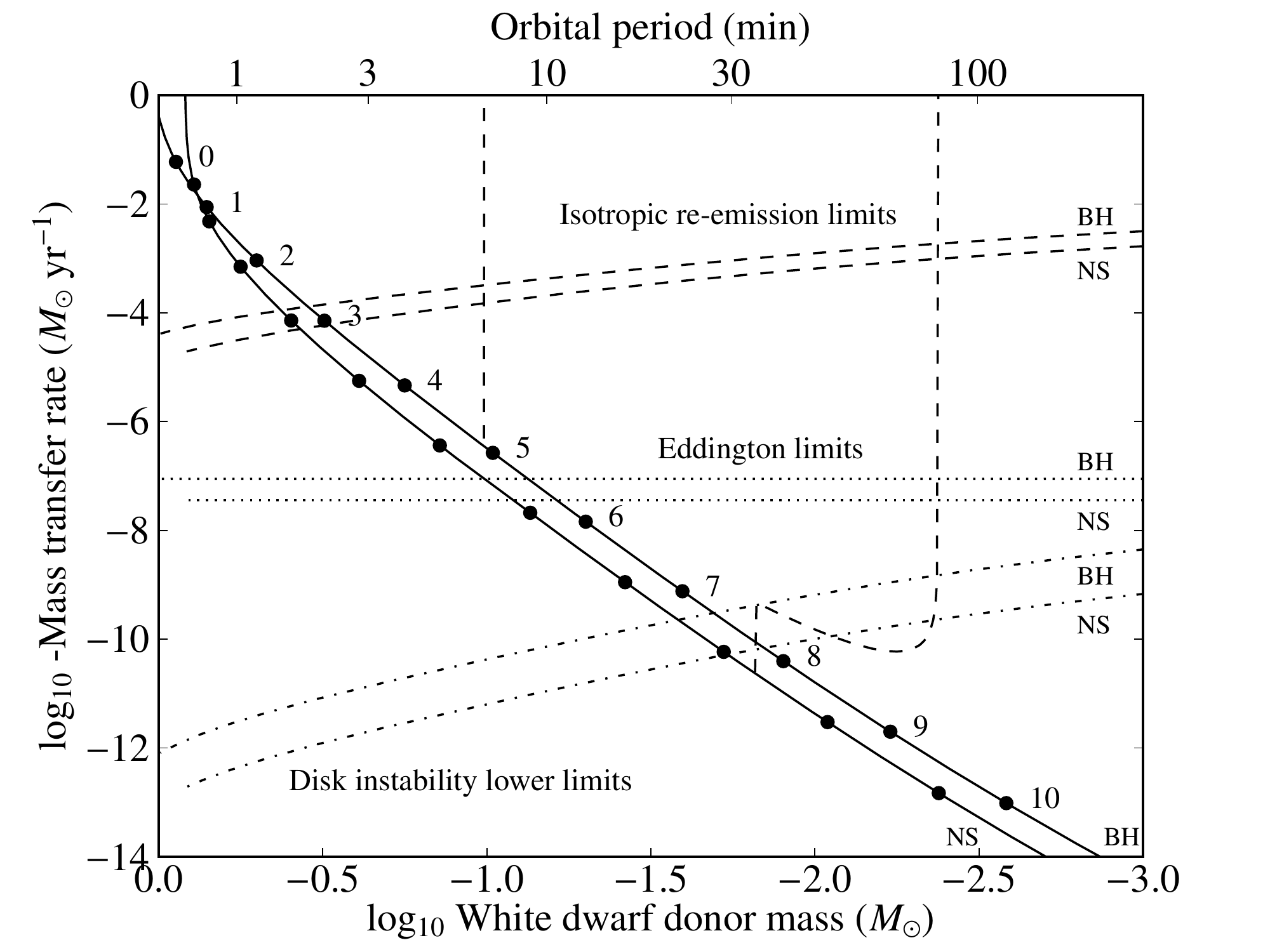}} 
\caption{Mass transfer rates for a helium white dwarf donor with an initially $1.4\ M_{\odot}$, $12\ \mbox{km}$ radius neutron star accretor (lower solid curve) and a $10\ M_{\odot}$ black hole accretor (upper solid curve), starting at the dynamical instability limit (i.e. ignoring possible merging), for full feedback. The dashed curves branching away from these represent the evolution assuming no feedback below $q = 0.01$ (Sect. \ref{reso}). The numbers correspond to the two circles on their left-hand side and represent the logarithm of the system age (\mbox{yr}), which is the time since the onset of mass transfer. Indicated are the isotropic re-emission regime (dashed, Eq. \ref{eqire}), the Eddington limit (dotted) and the accretion disk instability limit by \citet{zand2007} (dash-dotted, Sect. \ref{dim}) Note that the dashed neutron star curve at its local maximum 'touching' the upper dash-dotted curve is a coincidence.}
\label{fig:wddonor}
\end{figure}

In Fig. \ref{fig:wddonor}, the mass transfer rate starts very high because of the massive donor and especially because of the short orbit. The first part of the tracks lies above the isotropic re-emission limit (dashed, Sect. \ref{iresect}), which means that a system cannot survive here, this part of the tracks is shown just for illustration. Isotropic re-emission of super-Eddington mass transfer is possible for donor mass below $0.37\ M_{\odot}$ and orbital period above $1.8\ \mbox{min}$. For lower donor mass ($0.08\ M_{\odot}$, with a corresponding orbital period of $8\ \mbox{min}$), the mass transfer rate decreases below the Eddington limit (dotted), and from here on conservative mass transfer is possible (Fig. \ref{fig:epsilon}). At later times, the mass transfer rate reaches the thermal-viscous disk instability rate (dash-dotted, Sect. \ref{dim}). The donor mass is about $0.02\ M_{\odot}$ at this point, and the orbital period $28\ \mbox{min}$ in the case of a helium disk. Systems with a longer period than this do no longer have stable accretion disks.
After $10\ \mbox{Gyr}$ of mass transfer, a system with a $1.4\ M_{\odot}$ neutron star accretor would have a donor mass of $4.2 \cdot 10^{-3}\ M_{\odot}$ and an orbital period of $85$ min. In the case of a $10\ M_{\odot}$ black hole accretor, the donor mass would be $2.6 \cdot 10^{-3}\ M_{\odot}$ and the orbital period $111$ min. (The $10\ \mbox{Gyr}$ given here, as well as ages given elsewhere in the paper, is the time measured after the onset of mass transfer to the neutron star or black hole.) These systems technically do not fall under the present (observational) definition of UCXBs (an orbital period $\lesssim 60$ min).
Gravitational wave radiation is stronger in systems with a more massive accretor, therefore the black hole track in Fig. \ref{fig:wddonor} lies above the neutron star track (to be explained in Sect. \ref{macc}).

\begin{figure}
\resizebox{\hsize}{!}{\includegraphics{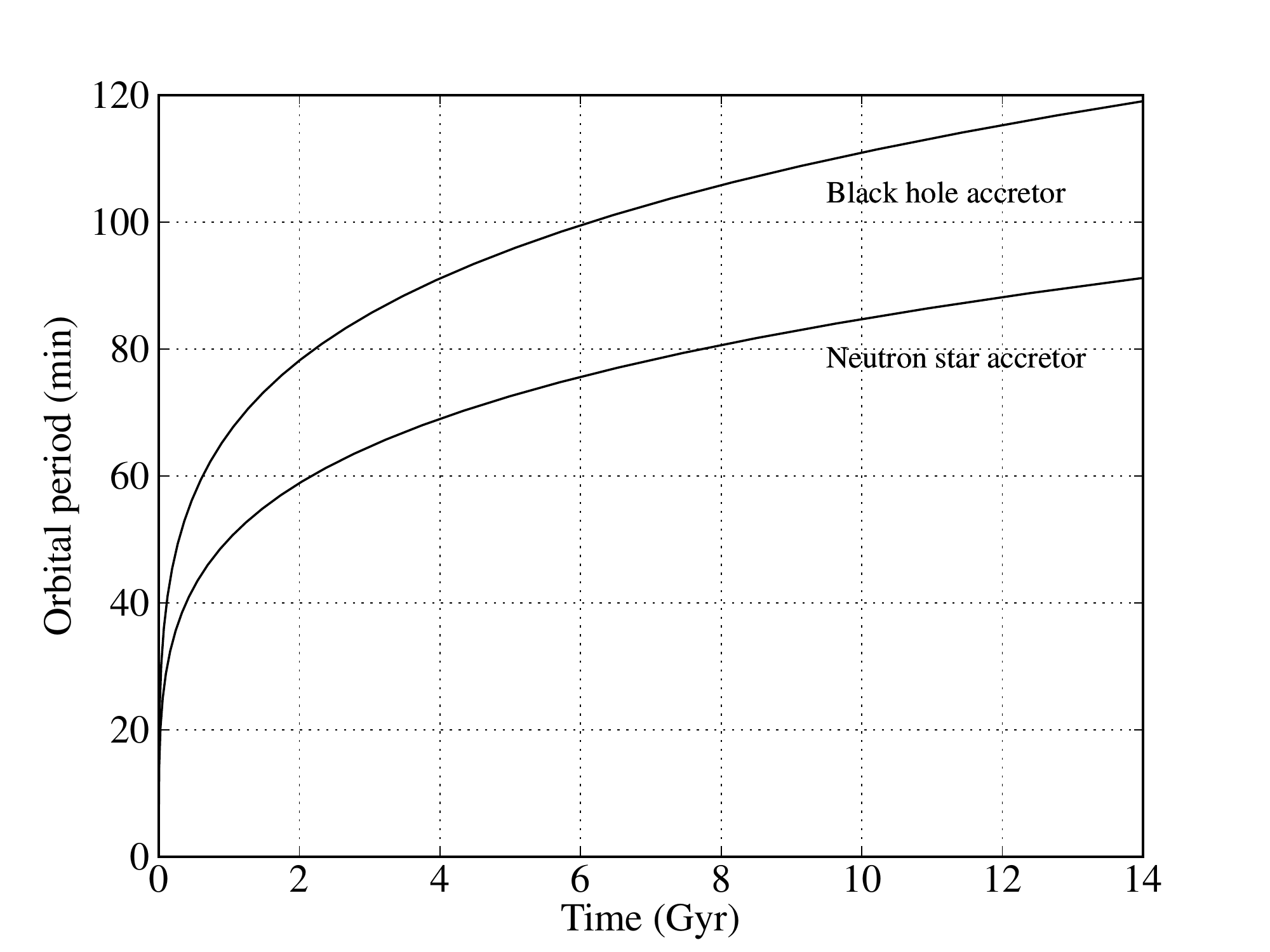}} 
\caption{Orbital period against time for a helium white dwarf donor with an initially $1.4\ M_{\odot}$, $12\ \mbox{km}$ radius neutron star (lower solid) and $10\ M_{\odot}$ black hole (upper solid) accretor.}
\label{fig:wdtime}
\end{figure}

Figure \ref{fig:wdtime} is a linear graph showing more clearly how the orbital period increases with time, initially the period increases very rapidly, and slowly later on. Systems with a massive accretor reach a given orbital period earlier on, because of their shorter evolutionary timescale.

\paragraph{The change in accretor mass during evolution}

For young systems with massive donors, the orbit is so compact that the donor is forced to lose mass at a high rate in order to avoid a merger. For these systems $\epsilon \approx 0$, because most mass is expelled from the system by isotropic re-emission. Once the transfer rate drops below the Eddington limit, $\epsilon = 1$ from Eq. (\ref{epsilon}) as all transferred mass can be accreted (Fig. \ref{fig:epsilon}). This happens at a donor mass of $0.08\ M_{\odot}$. Because $\epsilon$ goes from negligible to $1$ quickly (in terms of donor mass), the total mass increase of a neutron star $\Delta M_\mathrm{a}$ will not exceed $-\int \epsilon \dot{M}_\mathrm{d}\, \mbox{d}t=0.08\ M_{\odot}$ by Eq. (\ref{noncons}). The angular momentum associated with $\Delta M_\mathrm{a}$ is relevant when considering spin up of the accretor, in Sect. \ref{propeller}.

\paragraph{Observed potential long-period UCXBs}

The binary millisecond pulsar \object{HETE J1900.1--2455} has an orbital period of $83.3$ min, but a donor much more massive than systems of the same period following the evolution described here. The donor is probably a brown dwarf of $0.016 - 0.07\ M_{\odot}$\citep{kaaret2006}.

Because of their shorter evolutionary timescale, white dwarf - black hole systems can potentially reach an orbital period of $110\ \mbox{min}$ within the age of the Universe, if the system is not disrupted at some low mass ratio and mass transfer starts relatively quickly after the Big Bang. There are $2$ low-mass X-ray binaries known with $90 - 115\ \mbox{min}$ periods, viz. \object{1E 1603.6+2600} at $111\ \mbox{min}$ \citep{morris1990} and \object{XTE 1748--361} at $97\ \mbox{min}$ \citep{bhattacharyya2006}.
Another $2$ low-mass X-ray binaries have periods just outside this range, \object{SAX J1808.4--3658} at $120\ \mbox{min}$ \citep{chakrabarty1998} and \object{GS 1826--238} at $126\ \mbox{min}$ \citep{homer1998}, although the last period is uncertain \citep{liu2007}.
All these systems exhibit type I X-ray bursts so they must harbor a neutron star. The donors are not expected to be cold white dwarfs since these long periods can not have been reached by evolution on the gravitational wave timescale.

\paragraph{Why is conservative mass transfer less stable than non-conservative mass transfer?}

$\zeta_\mathrm{L}$ for conservative mass transfer ($\epsilon = 1$) is higher than for fully non-conservative mass transfer ($\epsilon = 0$) (the lower two dotted curves in Fig. \ref{fig:zetas}) implying dynamically less stable mass transfer for $\epsilon = 1$ by Eq. (\ref{mt2}), when we assume isotropic re-emission and full feedback of angular momentum from disk to orbit ($\eta = 0$). This may seem counter-intuitive since less mass ejection makes a system \emph{more} stable because less angular momentum is lost. The explanation is that the latter effect (represented by the $\gamma$-term in $\zeta_{a}$, Eq. (\ref{zetaa2})) is outweighed by the effect of the $(1-\beta)q$ ($= \epsilon q$ here) term for all $\epsilon$ and $q$ due to the unfavorable effect on the accretor mass: in the case of non-conservative mass transfer the accretor gains less mass, implying $a$ must increase more to compensate for this ($J_\mathrm{orb}^{2} = GaM_\mathrm{a}^{2}M_\mathrm{d}^{2}/M_\mathrm{tot}$) than in the conservative case. This effect is enhanced by the increase of $\zeta_\mathrm{r_\mathrm{L}}$ with $\epsilon$ (Eq. \ref{zetae}). For $\epsilon = 1$, the mass ratio $q = M_\mathrm{d}/M_\mathrm{a}$ decreases faster than for $\epsilon = 0$. The more extreme $q$ causes the donor Roche lobe to shrink more (or expand less) upon mass loss \emph{relative to $a$}.

\subsubsection{Why do massive accretors speed up evolution?}
\label{macc}

Figures \ref{fig:wdtracks} and \ref{fig:wddonor} show that UCXBs evolve faster if they have a more massive accretor. This is because evolution is driven by gravitational wave radiation. Here we give the explanation and look into some consequences.

\begin{itemize}

\item The effect of accretor mass on mass transfer rate:

For $q \lesssim 1$, the orbital period of a Roche-lobe filling system is only weakly dependent on the accretor mass (which can be seen from the Roche-lobe approximation by \citet{paczynski1971}), so the semi-major axis $a \propto M_\mathrm{tot}^{1/3}$ for fixed donor mass, according to Kepler's third law. From Eq. (\ref{mt2}) and $(\dot{J}/J)_\mathrm{\textsc{gwr}} = -(32/5) (G^{3}/c^{5}) M_\mathrm{d}M_\mathrm{a}M_\mathrm{tot}/a^{4}$ \citep{landau1975} follows that $\dot{M}_\mathrm{d} \propto M_\mathrm{a}/M_\mathrm{tot}^{1/3}$, so the mass transfer rate increases with accretor mass for a fixed donor, even though the orbital separation increases as well. This explains why the black hole track in Fig. \ref{fig:wddonor} lies above the neutron star track. The correction from $\zeta_\mathrm{d} - \zeta_\mathrm{L}$ is small unless $\zeta_\mathrm{d} - \zeta_\mathrm{L}$ approaches zero, as evidenced by the intersection of the tracks in Fig. \ref{fig:wddonor}, where the black hole system is still further away from its dynamical stability limit ($1.04\ M_{\odot}$). In reality, this intersection is not important since systems with an orbital period below $\sim 2\ \mbox{min}$ cannot survive.

\item Systems of the same age after the onset of mass transfer:

When we compare systems of the same age, mass transfer is \emph{lower} in the case of a more massive accretor (again, except when they are extremely young). The influence of the lower donor mass outweighs the influence of the more massive accretor in the gravitational wave equation, causing a lower mass transfer rate for more massive accretors of a fixed age. An old $1.4\ M_{\odot}$ neutron star accretor system has a mass transfer rate that is $\sim 50\%$ higher than a $10\ M_{\odot}$ black hole accretor system of the same age, as shown by the fits in Appendix \ref{wdfits}. (If the systems did not start with identical donors, the more massive donor will lose mass fast enough to (almost) catch up with the lower mass donor, on a timescale much shorter than the typical age of the systems.) A system with a lower accretor mass consistently experiences a higher mass transfer rate, but still its donor remains more massive. It can never catch up, because its mass transfer rate would drop below that of the black hole system in the case of nearly equal donor masses. Lastly, the gravitational wave timescale for a system with a neutron star accretor is $\sim 7\%$ longer than that of a $10\ M_{\odot}$ accretor system of the same age, and the orbital period is $\sim 25 \%$ shorter.

To summarize, in the case of an identical donor, $|\dot{M}_\mathrm{d}/M_\mathrm{d}|$ and $|\dot{M}_\mathrm{d}|$ are higher for systems with a higher accretor mass. For systems of the same age, however, $M_\mathrm{d}$ and $|\dot{M}_\mathrm{d}|$ are lower for a higher accretor mass, whereas $|\dot{M}_\mathrm{d}/M_\mathrm{d}|$ is marginally higher (i.e. mass is transferred on a slightly shorter timescale).

\item Consequences for AM CVn systems:

The above also implies that UCXBs evolve on a shorter timescale than AM CVn systems (which have white dwarfs both as donor and accretor) with an identical donor. The higher value of $\epsilon$ in AM CVns makes the mass transfer dynamically less stable and the mass transfer rate higher, but the accretor mass has a larger impact, unless the donor mass is close to the dynamical stability limit. For a low donor mass, $\zeta_\mathrm{d} - \zeta_\mathrm{L}$ is almost independent on the mass ratio and $\epsilon$, so the mass transfer rate only depends on the accretor mass. Also, AM CVns are expected to experience a higher mass transfer rate than UCXBs of the same age after onset of mass transfer, except again for very young systems. Old AM CVns have a more massive donor than equally old UCXBs.

\end{itemize}

\subsubsection{Evolution in the case of no feedback at low mass ratio}
\label{res_inst}

In Sect. \ref{explain_feedback} we mentioned several proposed mechanisms for reducing feedback of angular momentum from the disk to the orbit, but we think that none of these will actually work (Sect. \ref{reso}). In the case that after all a mechanism exists that leads to a reduction in or absence of feedback in systems below a certain low mass ratio,
here we briefly discuss the evolution of such systems, including the dynamical instability that may follow. We consider the case in which all feedback stops once $q < 0.01$, i.e. $\eta = 1$ and $\beta = 0$ in Eqs. (\ref{zetaa2}) and (\ref{zetaq2}) for these mass ratios.

Under this condition, the evolution of $\zeta$ for a neutron star accretor UCXB is shown by the upper solid curve in Fig. \ref{fig:zetas}. The sudden jump in $\zeta$ leads to an increased mass transfer rate per Eq. (\ref{mt2}), which speeds up evolution, therefore the low mass ratio where $\zeta_\mathrm{d}$ and $\zeta_\mathrm{L}$ meet can be reached well within the age of the Universe. Two evolutionary tracks are shown as the dashed curves branching away at $q = 0.01$ from the solid curves in Fig. \ref{fig:wddonor}.

\paragraph{Dynamically stable parameter ranges}

\begin{figure}
\resizebox{\hsize}{!}{\includegraphics{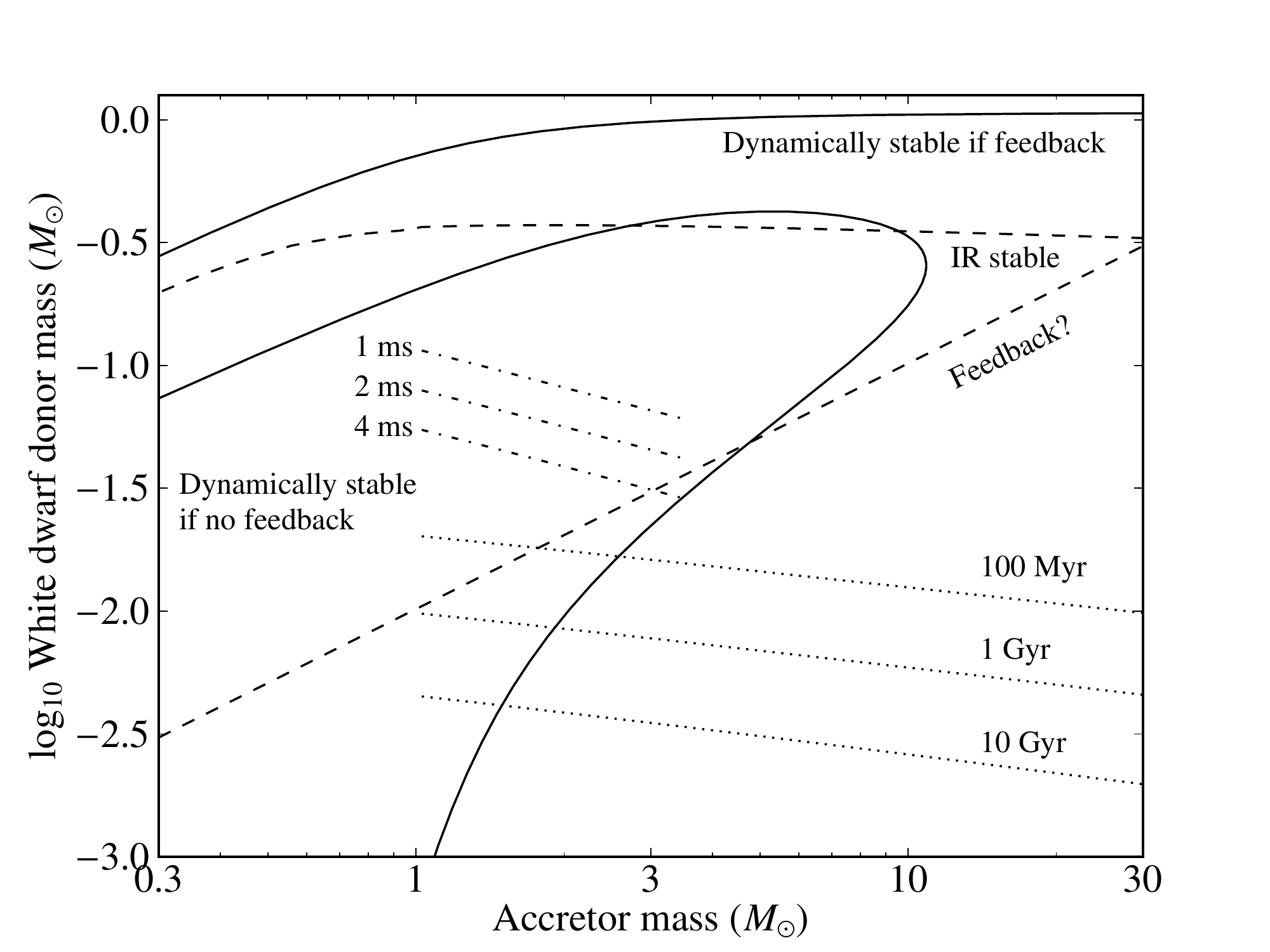}} 
\caption{Stability regimes in terms of zero-temperature helium white dwarf donor mass versus accretor mass. The upper solid curve is the upper limit for dynamical stability in the presence of an accretion disk that feeds back angular momentum, where almost all transferred matter is lost via isotropic re-emission. The upper dashed curve is the upper stability limit for isotropic re-emission and the lower dashed curve indicates the proposed $q = 0.01$ lower mass ratio limit for feedback (Sect. \ref{reso}). The lower solid curve encloses the parameter space where systems are dynamically stable in the absence of angular momentum feedback; all matter enters and stays in the disk. The dash-dotted lines indicate when the propeller effect comes into play according to Eq. (\ref{mdotprop}) for several neutron star spin periods and an equatorial magnetic field of $10^{8.5}$ G. The dotted isochrones indicate the donor masses for system ages (since the onset of mass transfer) of $0.1, 1, 10\ \mbox{Gyr}$ \emph{based on full feedback}.
The dynamical stability and isotropic re-emission rates are independent of accretor radius and type, hence these curves are continuous. Moreover, the dynamical stability rates are independent of accretor type and therefore also hold for white dwarf accretors.}
\label{fig:shark}
\end{figure}

Figure \ref{fig:shark} shows which combinations of donor and accretor masses are dynamically stable, both in the case of feedback and the case of no feedback. The accretor mass has a large influence on when an instability occurs, if at all. The bottom left region enclosed by the solid line is dynamically stable for mass transfer, even in the absence of feedback. Initial mass transfer is dynamically unstable for every $M_\mathrm{d}$ if $M_\mathrm{a} > 11\ M_{\odot}$, because of high specific angular momentum of the stream in low mass ratio systems.
The upper border of this region shows that the highest donor mass that can possibly avoid dynamical instability in the case of no feedback is $0.42\ M_{\odot}$, for an accretor mass of $5.3\ M_{\odot}$.\footnote{This upper limit is a result of the $\zeta_\mathrm{L}$ function in Eq. (\ref{zetarl}) with $\beta = \epsilon = 0$ having a minimum of $\zeta_\mathrm{L}=-0.48$ at $q=0.080$, and $\zeta_\mathrm{d} = -0.48$ is solved by $M_\mathrm{d} = 0.42\ M_{\odot}$ in Eq. (\ref{zetawd}).} As explained in Sect. \ref{deciding}, in practice the upper border will be replaced by a limit of $\sim 0.5\ M_{\odot}$.
Note that the donor mass at which the low mass ratio dynamical instability could occur is very sensitive to accretor mass near $M_\mathrm{a} = 1.4\ M_{\odot}$. At exactly which donor mass this instability occurs depends on the mass-radius relation of the donor, which in turn depends on its temperature and composition.

The upper limit for dynamical stability in the case of full feedback is represented by the upper solid curve in Fig. \ref{fig:shark}. Because $\zeta_\mathrm{L}$ (Eq. \ref{zetarl}) has a minimum of $-5/3$ for $q \to 0$ regardless of $\epsilon$ and $\gamma$, any system with a zero-temperature white dwarf donor mass exceeding $1.07\ M_{\odot}$ is dynamically unstable, even for arbitrarily high accretor mass.\footnote{Around an intermediate mass or supermassive black hole, an accretion disk can only exist if $M_\mathrm{a} \lesssim 10^{5}\ M_{\odot}$ in the case of a $0.6\ M_{\odot}$ white dwarf donor and $M_\mathrm{a} \lesssim 10^{6}\ M_{\odot}$ in the case of a $0.1\ M_{\odot}$ white dwarf donor, since otherwise the innermost stable circular orbit exceeds the semi-major axis.}

\paragraph{Age of disruption}

If feedback would suddenly stop at $q = 0.01$, a dynamical instability would happen immediately if $M_\mathrm{a} > 4.7\ M_{\odot}$, and later for less massive accretors (shown by the intersection between the lower solid and dashed curves in Fig. \ref{fig:shark}).
For an UCXB with a $1.4\ M_{\odot}$ accretor, the instability would happen at an age of $300\ \mbox{Myr}$, for $M_\mathrm{d} = 3.2 \cdot 10^{-3}\ M_{\odot}$ (the upper-left intersection at $q = 2.3 \cdot 10^{-3}$ in Fig. \ref{fig:zetas}).
The reason such a low donor mass could be reached well within $10\ \mbox{Gyr}$ is the increased mass transfer rate caused by the extra angular momentum sink, about one order of magnitude for a $1.4\ M_{\odot}$ accretor (Fig. \ref{fig:wddonor}).
A $10\ M_{\odot}$ black hole system would reach $q = 0.01$ after already $80$ kyr.

Disruption of low-mass donors would happen later than $10\ \mbox{Gyr}$ if $M_\mathrm{a} < 0.6\ M_{\odot}$, this could apply to AM CVn systems. UCXBs with any realistic accretor mass would suffer disruption well within $10\ \mbox{Gyr}$.

\subsection{System evolution during the propeller phase}
\label{propeller}

Even though most likely no dynamical instability occurs at low mass ratios, the regular accretion rates as described in Sect. \ref{acc} are idealized. Here we investigate the UCXB evolution taking a magnetized neutron star accretor into account, using angular momentum and energy considerations. This gives insight into the feasibility of the propeller effect (Sect. \ref{prop}) removing matter from the system and hence suppressing accretion onto and X-ray emission from UCXBs.

\subsubsection{Steady mass transfer}

First we assume the absence of disk instability (Sect. \ref{dim}). This means that matter approaches the magnetosphere at a constant rate (that is, only varying on the evolutionary timescale).
From a magnetic perspective, the evolution can be divided in a spin up and a spin down stage.

\paragraph{Spin up}

Initially the mass transfer rate is so high that the kinetic energy in the disk completely dominates the magnetic field energy; the Alfv\'en radius lies either at the surface or just outside the neutron star. Moreover, the relatively low initial accretor spin frequency (and correspondingly large corotation radius) ensures that any interaction between disk and magnetic field that is still present, causes the neutron star to spin up rather than matter to be expelled, because the fast rotating disk applies a torque on the field lines, which leads to transfer of angular momentum from disk to neutron star. We assume that as long as the Alfv\'en radius is smaller than $2^{1/3} \approx 1.26$ times the corotation radius (Sect. \ref{critmt}), all arriving matter is accreted, and the amount of angular momentum added to the accretor corresponds to the specific angular momentum of disk matter at the Alfv\'en radius. Specific angular momentum of matter in a Keplerian orbit is proportional to the square root of the distance to the accretor, and we will find that the typical ratio of the outer disk radius to the Alfv\'en radius exceeds $10^{3}$ during the spin-up stage. Therefore, the fraction of angular momentum that is not transferred back via the outer disk is at most a few percent, and our initial assumption of dynamically stable mass transfer holds.

\begin{figure}
\resizebox{\hsize}{!}{\includegraphics{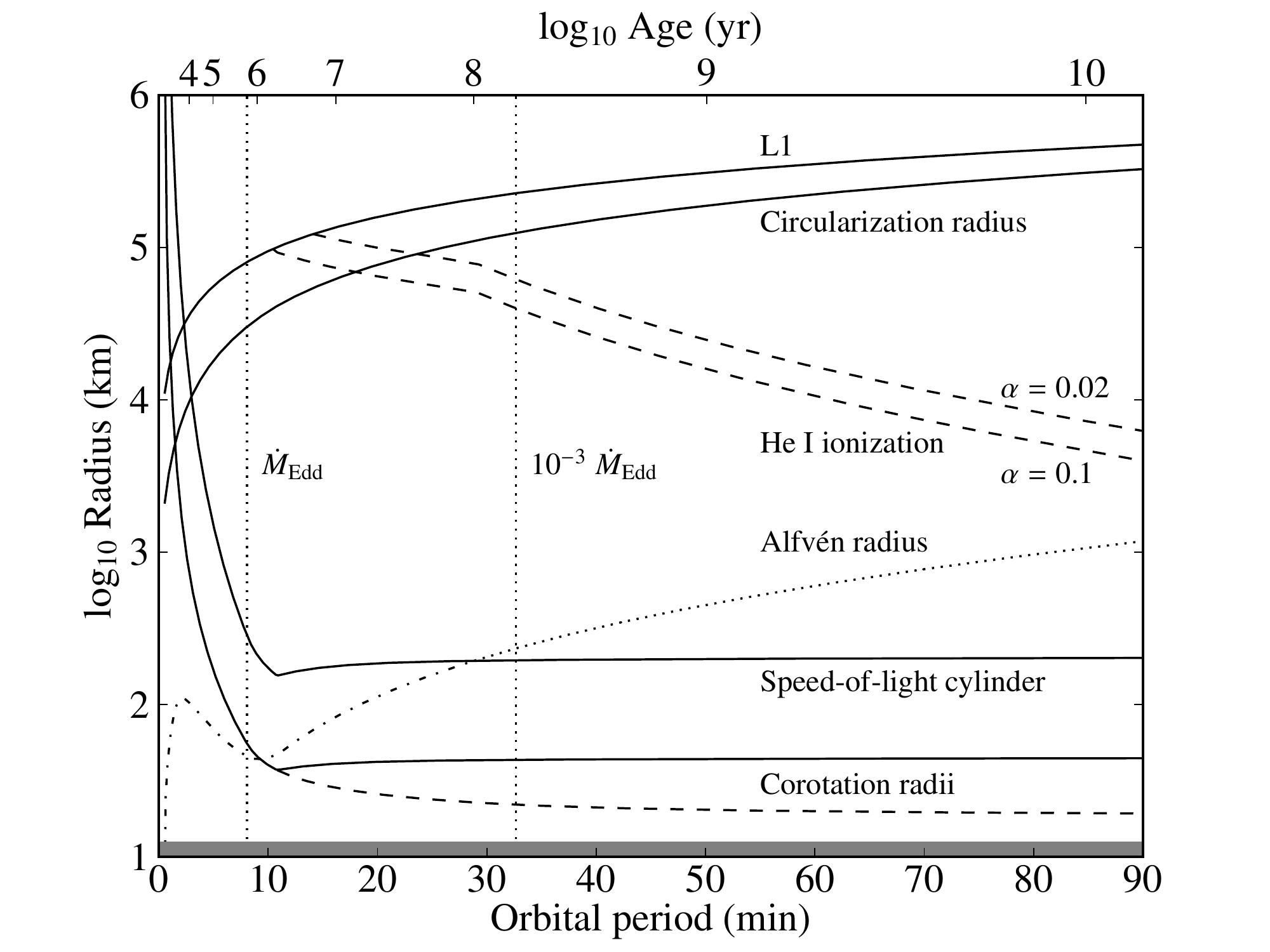}} 
\caption{Radii in UCXB accretion disk versus orbital period for a $1.4\ M_{\odot}$ neutron star accretor with a residual equatorial magnetic field of $10^{8.5}$ G. The shaded layer at the bottom represents the neutron star surface, at a radius of $12$ km. The curve annotated L1 shows the distance of the first Lagrangian point to the neutron star, which is an upper limit on the outer disk radius. The circularization radius by \citet{verbunt1988} is discussed in Sect. \ref{reso}. The two upper dashed curves indicate the radii where the temperature of a helium disk \citep{dunkel2006} equals the He I ionization temperature, for two values of the viscosity parameter $\alpha$, as discussed in Sect. \ref{dim_intro}. The solid corotation radius curve applies to the full propeller case, the dashed one to the hypothetical full accretion case. The speed-of-light cylinder radius and Alfv\'en radius are also shown.}
\label{fig:prop_radii_per}
\end{figure}

As the neutron star spins up, the corotation radius approaches the neutron star. At the same time the Alfv\'en radius moves outwards as a result of decreasing mass transfer. At an orbital period of $10$ min (and a system age of $1$ Myr) these two meet. The surface magnetic field strength at this time is determined by how much mass the accretor has gained up to this point (Sect. \ref{acc}). We adopt the empirical relation for the accretion-induced field decay by \citet{shibazaki1989}, with a residual field strength of $B_\mathrm{NS} = 10^{8.5}$ G \citep{zhang2006}.
We also assume that the neutron star does not spin when the mass transfer starts, i.e. $J_\mathrm{NS} = 0$, since the earlier common-envelope stage is too short-lived to allow for significant accretion. Also, the strong initial magnetic field suggests a very low initial spin frequency relative to the eventual spin frequency.

The amount of angular momentum added to the neutron star by accretion is

\begin{equation}
    \label{addedam}
    \mathrm{d} J_\mathrm{NS} = \mathrm{d} M_\mathrm{NS} \sqrt{G M_\mathrm{NS} R_\mathrm{in}}
\end{equation}
where $R_\mathrm{in}$ is the inner disk radius, which determines the specific angular momentum of matter in a Keplerian orbit at this radius. We assume that matter stops losing angular momentum outwards once it moves inside this radius. $R_\mathrm{in} = R_{\mu}$ \citep{li1999} except when the magnetic field does not dominate the Kepler flow anywhere, in which case the stellar surface is taken, or when $R_{\mu}$ lies outside the speed-of-light cylinder radius $R_{c} = c/\omega_\mathrm{s}$, where corotating field lines would travel at the speed of light. There, field lines are open and particles can escape by following them outwards. The value of the inner radius is summarized as

\begin{equation}
    \label{racc}
    R_\mathrm{in} = \min\,(R_{c},\max\,(R_{\mu},R_\mathrm{NS})).
\end{equation}
From $J_\mathrm{NS}$ the spin angular frequency $\omega_\mathrm{s}$ is found using $J_\mathrm{NS} = I_\mathrm{NS} \omega_\mathrm{s}$, where $I_\mathrm{NS} \approx 0.357 M_\mathrm{NS} R_\mathrm{NS}^{2}$ is the moment of inertia of a $1.4\ M_{\odot}$, $12$ km neutron star \citep{lattimer2005}. We consider angular momentum accretion rather than energy accretion, because the latter is complicated by radiation losses and orbit coupling.

\paragraph{Spin down}

Once the Alfv\'en radius moves beyond the corotation radius, the torque is reversed, and matter in Keplerian orbits between the corotation radius and the Alfv\'en radius may be accelerated by the super-Keplerian velocity of the field lines, reducing the accretion rate. Matter near $R_{\mu}$ (or $R_{c}$, once smaller) that reaches the escape velocity then leaves the system on trajectories outside the orbital plane, the angle between the outflow and the plane being smaller for a stronger propeller effect \citep{lovelace1999,ustyugova2006}.

Due to its accretion history, the neutron star has gained a large amount of rotational kinetic energy at this stage. The evolution of spin frequency $\omega_\mathrm{s}$ once $R_\mathrm{\mu} > 2^{1/3} R_\mathrm{co}$ depends on whether there is enough energy, and whether it can be employed to unbind all or some fraction of arriving matter. At this point we assume that there is sufficient energy to unbind all transferred matter when $R_\mathrm{\mu} > 2^{1/3} R_\mathrm{co}$, and will later find that this is indeed the case. Then, the energy of the neutron star changes as

\begin{equation}
    \label{erot}
    \mathrm{d} E_\mathrm{NS} = \frac{G M_\mathrm{NS} \dot{M}_\mathrm{d}}{2 R_\mathrm{in}} \mathrm {d}t,
\end{equation}
where $\mathrm{d} E_\mathrm{NS}$ and $\dot{M}_\mathrm{d}$ are negative, and the factor $2$ appears because the total energy in a Keplerian orbit is half the potential energy. The rotational kinetic energy of the neutron star is $E_\mathrm{NS} = \frac{1}{2} I_\mathrm{NS} \omega_\mathrm{s}^{2}$, which yields $\omega_\mathrm{s}$.\footnote{The solution for $\omega_\mathrm{s}$ is implicit since $\omega_\mathrm{s}$ depends on $E_\mathrm{NS}$, which itself depends on the speed-of-light cylinder radius $R_{c}$ via Eqs. (\ref{erot}) and (\ref{racc}).}

\paragraph{Spin evolution}

Figure \ref{fig:prop_radii_per} illustrates the evolution of the accretion disk, magnetosphere and corotation radius for an UCXB. The dotted part of the Alfv\'en radius curve (Eq. \ref{alfven}) lies beyond the speed-of-light cylinder radius, and therefore the field lines cannot corotate in this region. The corotation radius is given by Eq. (\ref{corot}) and is proportional to $P_\mathrm{s}^{2/3}$, where the accretor spin period $P_\mathrm{s} = 2\pi/\omega_\mathrm{s}$. The solid corotation curve shows the evolution when no matter is accreted once $R_{\mu} > 2^{1/3} R_\mathrm{co}$, but instead all transferred matter is unbound from radius $R_\mathrm{in}$, extracting energy from the rotation of the neutron star. For comparison, the dashed corotation curve shows the spin evolution in the hypothetical case in which, once $R_{\mu} > 2^{1/3} R_\mathrm{co}$, all matter would continue to be accreted with specific angular momentum corresponding to $R_\mathrm{in}$ instead of stopped by the magnetosphere. Before $R_{\mu} = 2^{1/3} R_\mathrm{co}$, both assume full accretion so there these two are equal.

Because the corotation radius stays within the Alfv\'en radius, apparently the early accretion at a high rate has sufficiently spun up the neutron star to provide enough energy to unbind any fraction of transferred matter during the whole remainder of the evolution, when the mass transfer rate is low. This justifies our assumption above Eq. (\ref{erot}).

By using a lower effective equatorial surface magnetic field strength to accommodate for disk accretion, as discussed in Sect. \ref{diskacc}, less angular momentum is accreted per unit mass due to the smaller Alfv\'en radius, but also accretion can continue for a longer time because the propeller effect appears at a later stage. The result is that the neutron star is spun up to a lower spin period, while subsequently less energy is needed to unbind arriving matter because the propeller effect starts at a lower mass transfer rate. The eventual spin period is lower in the case of a lower effective magnetic field.

\begin{figure}
\resizebox{\hsize}{!}{\includegraphics{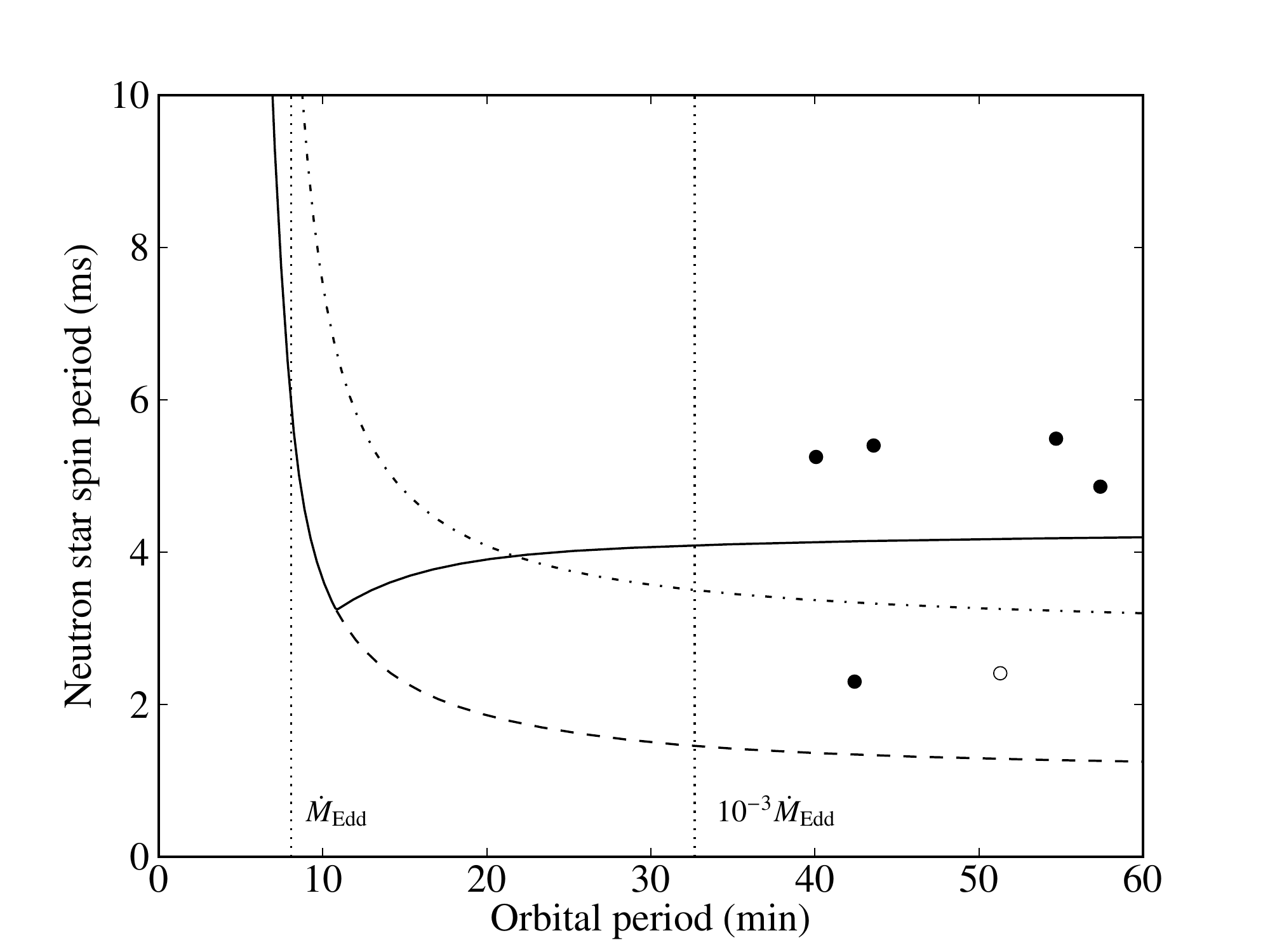}} 
\caption{Spin period versus orbital period for a $1.4\ M_{\odot}$ neutron star accretor with a residual equatorial magnetic field of $10^{8.5}$ G, assuming all transferred matter is propelled out of the system once $R_{\mu} > 2^{1/3} R_\mathrm{co}$ (solid, Sect. \ref{critmt}), or hypothetically keeps being accreted (dashed), or hypothetically is accreted at the neutron star surface all the time (dash-dotted). Observed millisecond pulsar-UCXB parameters are represented by filled disks; \object{XTE J1807--294} at an orbital period of $40.07$ min \citep{markwardt2003b} and a spin period of $5.25$ ms \citep{markwardt2003}, \object{XTE J1751--305} at $42.42$ min, $2.30$ ms \citep{markwardt2002}, \object{XTE J0929--314} at $43.58$ min, $5.40$ ms \citep{galloway2002}, \object{SWIFT J1756.9--2508} at $54.70$ min, $5.49$ ms \citep{krimm2007} and \object{NGC 6440 X-2} at $57.3$ min, $4.86$ ms \citep{altamirano2010}. The open disk represents \object{4U 0614+091}, which harbors a $2.41$ ms pulsar \citep{strohmayer2008} and has a tentative orbital period of $51.3$ min \citep{shahbaz2008,hakala2011}.}
\label{fig:prop_pspin}
\end{figure}

Figure \ref{fig:prop_pspin} shows the spin period directly, rather than the related corotation radius. The hypothetical case in which the propeller effect does not work, so all matter is accreted at the stellar surface carrying the local Keplerian specific angular momentum, is shown by the dash-dotted curve (Sect. \ref{diskacc}). For comparison we include the observed UCXB neutron star spin periods, which are too small in number and have too scattered spin periods to have any constraining value to our model.

Now we know that full propeller is energetically possible for UCXBs, does it actually happen? The observed neutron star spin periods, typically $5$ ms, suggest that at least some spin down is plausible. On the other hand, the faint transient X-ray binary \object{SAX J1808.4--3658} proves that at least some accretion is still possible onto a very fast spinning ($2.5$ ms) neutron star, since it shows type I X-ray bursts \citep{zand1998}. \citet{spruit1993,rappaport2004} showed that some accretion likely persists assuming the inner disk radius comes quite close to the corotation radius. They also mentioned that this disk structure probably breaks down at very low mass transfer rate, and that full propelling probably happens then.
Simulations by \citet{dangelo2011} showed that accelerated matter does not necessarily leave the system, because the inner disk radius may remain trapped at some short distance outside the corotation radius. This scenario, however, appears more likely for stronger magnetic fields ($\sim 10^{12}$ G) than the ones considered here, and for stars with a much shorter spin-down timescale than millisecond pulsars. Therefore, a non-accreting state seems more likely for UCXBs containing a recycled neutron star, and the propeller effect may work depending on whether the inner disk radius exceeds the corotation radius by a sufficiently large ratio.

Whether accretion at very low mass transfer rate is possible or not in a system with a stable accretion disk perhaps is not the right question to ask, since accretion most likely occurs irregularly. Figure \ref{fig:prop_radii_per} suggests that a helium disk (and also a carbon-oxygen disk, which is not shown) remains divided in a neutral outer part and an ionized inner part, where the inner part at all times is too large for the magnetosphere to disrupt. This implies that the disk instability remains relevant and will come into play once the mass transfer rate becomes too low (Sect. \ref{dim_intro}).

\subsubsection{Non-steady mass transfer}

Varying mass flow rates and densities near the accretor have a large impact on the interaction between the disk and the magnetic field. For instance, \citet{spruit1993} described cycles of accretion, during which matter piles up outside the Alfv\'en radius until the local density is such that the inner disk forces the Alfv\'en radius inwards, allowing accretion. When the disk density decreases, the Alfv\'en radius moves outwards again.
In Fig. \ref{fig:prop_radii_per} it can be seen that the Alfv\'en radius at Eddington-limit mass transfer is almost equal to the late-time corotation radius, but taking into account the decreasing magnetic field, the Alfv\'en radius will always lie well inside the corotation radius at the maximum accretion rate, so a near-Eddington outburst can always be accreted.

\begin{figure}
\resizebox{\hsize}{!}{\includegraphics{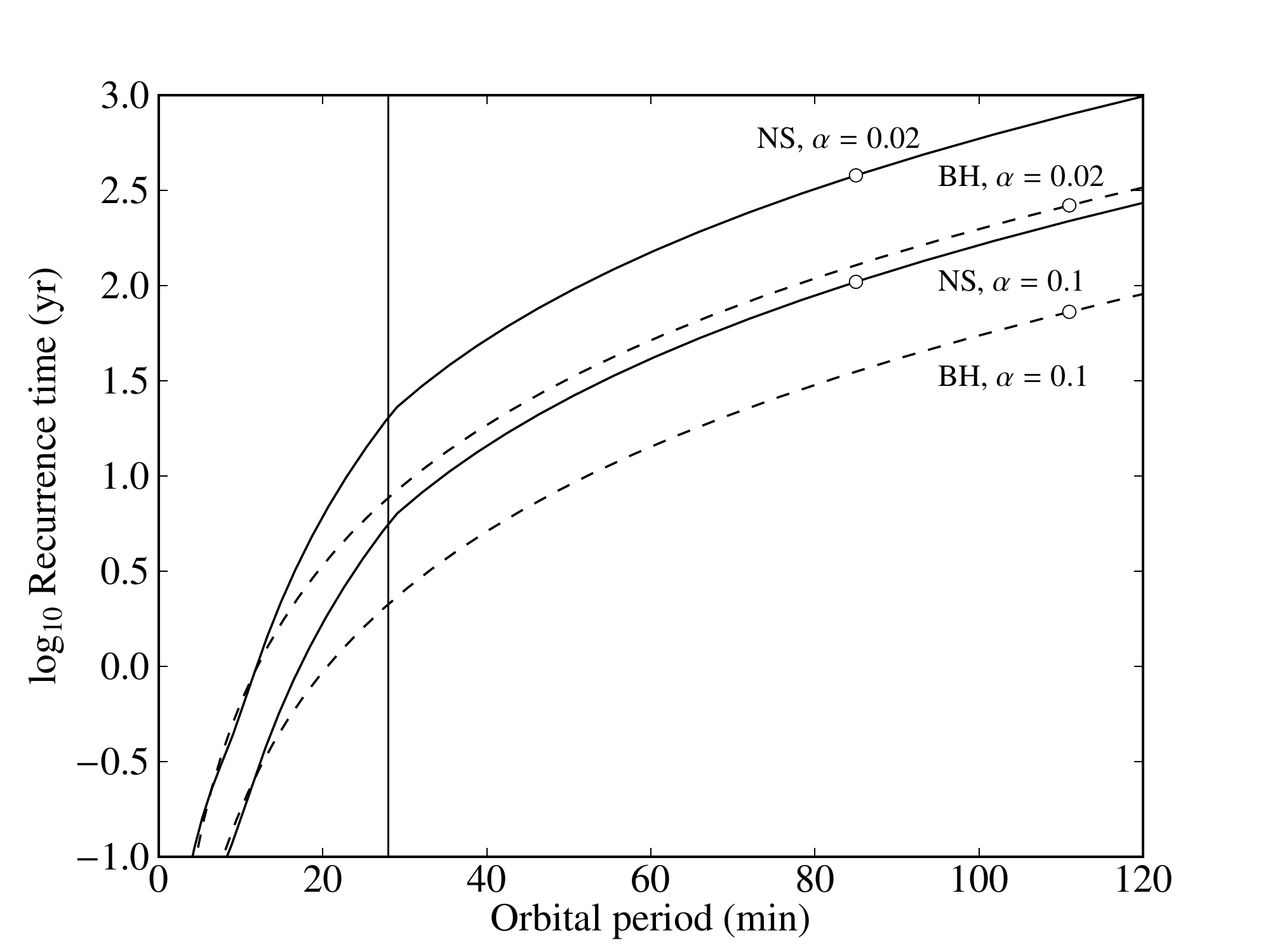}} 
\caption{Recurrence time for a $1.4\ M_{\odot}$ neutron star accretor with a residual equatorial magnetic field of $10^{8.5}$ G (solid) and for a $10\ M_{\odot}$ black hole accretor (dashed) with a helium disk described by \citet{dunkel2006}, for two values of the viscosity parameter $\alpha$, of which $0.02$ probably best describes the disk viscosity in quiescence. Disk instability becomes relevant to the right of the vertical line at $28$ min. The circles indicate an age of $10$ Gyr. The discontinuity in the slopes of the neutron star curves at an orbital period of $29$ min is because the speed-of-light cylinder defines the inner disk radius beyond this period (Fig. \ref{fig:prop_radii_per}).}
\label{fig:rectime}
\end{figure}

In Sect. \ref{dim} we have discussed under which circumstances accretion could become irregular due to thermal-viscous disk instability. When the mass transfer rate has become sufficiently low, the disk periodically collapses and builds, which leads to high mass transfer rates near the accretor every cycle. The recurrence time is the time it takes until the empty disk is refilled. Figure \ref{fig:rectime} shows the recurrence time for our UCXB model. The inner disk radius $R_\mathrm{in}$ is given by Eq. (\ref{racc}) in the case of a neutron star accretor and $3$ Schwarzschild radii in the case of a black hole accretor. The outer radius is estimated by $R_\mathrm{out} \approx 0.9 d_\mathrm{L1}$ based on Fig. \ref{fig:radii}. The recurrence time is increasing because the mass transfer rate decreases faster than the disk mass. During quiescence the disk viscosity is low, of the order of $\alpha = 0.02$ \citep{lasota2001}. Systems of age $10^{9-10}$ yr with a neutron star accretor have a recurrence time of hundreds of years. In the case of an equally old system with black hole accretor, this time is a bit shorter. The long recurrence time suggests that systems are only rarely observable as bright X-ray sources, so that at any given time, only a small fraction of a population is visible.

\section{Discussion and conclusions}
\label{disc}

We have studied the evolution of ultracompact X-ray binaries, and in particular their characteristics at an old age, when they reach a very low mass ratio and low mass transfer rate. The aim is to investigate how likely the occurrence of a dynamical instability is, and how important the propeller effect and disk instability are.

If the feedback of angular momentum from disk to orbit were to be reduced, the lifetime of an UCXB would be dramatically shortened. Based on accretion disk simulations (Sect. \ref{reso}), we conclude that an accretion disk in a system of any realistically achievable mass ratio is neither significantly affected by resonant orbits, nor does it expand outside the accretor Roche lobe. The latter is in agreement with \citet{priedhorsky1988}. Most likely the disk is still capable of full feedback of angular momentum to the orbit so that the donor will never be disrupted. Consequently, unless a new mechanism is discovered that limits or stops feedback, or pulsar radiation evaporates the donor by heating and stripping off the outer layers \citep{kluzniak1988,ruderman1989,shaham1992}, or UCXBs become temporarily detached at some stage, the vast majority of UCXBs today will have orbital periods around $70 - 80$ min and experience low mass transfer rates of $10^{-13} - 10^{-12}\ M_{\odot} \mbox{yr}^{-1}$, depending on the precise mass-radius relation of the donor. (Systems that started mass transfer more recently will have shorter orbital periods, but since the evolutionary timescale increases rapidly with increasing orbital period, most systems are expected to have a long orbital period at the present. The time between the zero-age main sequence and the onset of mass transfer is in general short, as will be shown in a forthcoming paper.)

However, all $\sim 15$ known UCXBs have much shorter orbital periods, below $60$ min. Since these short-period systems evolve on a much shorter timescale than long-period systems, there must exist a much larger population of apparently invisible, old, long-period UCXBs in the Galaxy. A strong selection effect hides the large population of relatively common systems so that we see only a couple of much rarer systems during their short-lived high mass transfer phase. The bolometric luminosity corresponding to a constant mass transfer rate of $\sim 10^{-13}\ M_{\odot} \mbox{yr}^{-1}$ is $\sim 5 \cdot 10^{32}\ \mbox{erg s}^{-1}$, within the observational limits of the best present-day X-ray telescopes even at a distance of $8$ kpc (the distance to the Galactic Bulge) if these systems emit predominantly in X-rays. Then, the only consistent resolutions are that these long-period systems are even fainter than expected, or visible only a small fraction of the time. These resolutions are supported by the propeller effect and the disk instability model, respectively.

The propeller effect probably strongly influences accretion in UCXBs, especially at very low mass transfer rates. In the event of a stable disk, rotational kinetic energy stored in the neutron star during the accreting stage is sufficient to expel all matter arriving at the magnetosphere from the system once the propeller effect starts to work. Therefore, systems with a stable accretion disk could have a significantly reduced X-ray luminosity.

However, the disk instability model, which describes how disks periodically build and collapse, still applies at the very low mass transfer rates expected in UCXBs. During a collapse, matter is capable of penetrating the magnetosphere and reach the stellar surface, which means that observationally the system appears as a transient X-ray source. Based on estimates of the disk structure, the recurrence time of old ($10$ Gyr) systems is of the order of a few $100$ yr, depending on at what mass transfer rate the disk becomes unstable, among other things.
Since black hole accretor systems do not suffer from the propeller effect, these would be expected to be better visible, but none have been discovered so far. Apparently they are much rarer than their neutron star accretor counterparts, or disk instability is the dominant process in making low mass ratio UCXBs nearly invisible.

A third mechanism that can inhibit the X-ray visibility of a black hole UCXB is the inefficiently radiating disk (ADAF -- advection dominated accretion flow) \citep{narayan1994,lasota2008}. In ADAFs, most of the thermal energy stored in the inner disk is accreted into the event horizon. For a neutron star UCXB this does not apply as the energy would still be released upon accretion.

Lastly, UCXBs with a low mass transfer rate may also become fainter in X-rays if they emit a significant fraction of their energy as UV or even optical radiation. The \citet{dunkel2006} central disk temperature for helium composition is given by $T_\mathrm{c} \propto \dot{M}^{3/10}$ or $\dot{M}^{2/5}$ depending on the choice of opacity, while a thick disk has $T \propto \dot{M}^{1/4}$, so the disk spectrum shifts to longer wavelengths as the mass transfer rate decreases. The hot inner part of the disk, origin of most radiation, still emits mainly X-rays, with the result that the disk as a whole emits more than half of its energy as X-ray radiation, even at the lowest mass transfer rates we find. The inner disk, however, is affected by the magnetic field in the case of neutron star accretors (in particular at low mass transfer rate), and can become an ADAF in the case of black hole accretors. In these cases the outer disk spectrum becomes more important and the system may emit most radiation in the UV, which is difficult to observe due to absorption \citep{campana2000}.

The consequences of the evolutionary behavior of UCXBs for the population in the Galactic Bulge will be the topic of a forthcoming paper.

\begin{acknowledgements}
We thank E. K\"{o}rding for helpful discussions on accretion disks. LMvH is supported by the Netherlands Organisation for Scientific Research (NWO). GN and RV are supported by NWO Vidi grant $016.093.305$.
\end{acknowledgements}

\bibliographystyle{aa}
\bibliography{lennart_refs}

\begin{thebibliography}{99}
\expandafter\ifx\csname natexlab\endcsname\relax\def\natexlab#1{#1}\fi

\bibitem[{{Alpar} {et~al.}(1982){Alpar}, {Cheng}, {Ruderman}, \&
  {Shaham}}]{alpar1982}
{Alpar}, M.~A., {Cheng}, A.~F., {Ruderman}, M.~A., \& {Shaham}, J. 1982, \nat,
  300, 728

\bibitem[{{Altamirano} {et~al.}(2010){Altamirano}, {Patruno}, {Heinke},
  {Markwardt}, {Strohmayer}, {Linares}, {Wijnands}, {van der Klis}, \&
  {Swank}}]{altamirano2010}
{Altamirano}, D., {Patruno}, A., {Heinke}, C.~O., {et~al.} 2010, \apjl, 712,
  L58

\bibitem[{{Aly} \& {Kuijpers}(1990)}]{aly1990}
{Aly}, J.~J. \& {Kuijpers}, J. 1990, \aap, 227, 473

\bibitem[{{Begelman}(1979)}]{begelman1979}
{Begelman}, M.~C. 1979, \mnras, 187, 237

\bibitem[{{Bhattacharya} \& {van den Heuvel}(1991)}]{bhattacharya1991}
{Bhattacharya}, D. \& {van den Heuvel}, E.~P.~J. 1991, \physrep, 203, 1

\bibitem[{{Bhattacharyya} {et~al.}(2006){Bhattacharyya}, {Strohmayer},
  {Markwardt}, \& {Swank}}]{bhattacharyya2006}
{Bhattacharyya}, S., {Strohmayer}, T.~E., {Markwardt}, C.~B., \& {Swank}, J.~H.
  2006, \apjl, 639, L31

\bibitem[{{Bildsten}(2002)}]{bildsten2002}
{Bildsten}, L. 2002, \apjl, 577, L27

\bibitem[{{Bonsema} \& {van den Heuvel}(1985)}]{bonsema1985}
{Bonsema}, P.~F.~J. \& {van den Heuvel}, E.~P.~J. 1985, \aap, 146, L3

\bibitem[{{Campana} \& {Stella}(2000)}]{campana2000}
{Campana}, S. \& {Stella}, L. 2000, \apj, 541, 849

\bibitem[{{Chakrabarty} \& {Morgan}(1998)}]{chakrabarty1998}
{Chakrabarty}, D. \& {Morgan}, E.~H. 1998, \nat, 394, 346

\bibitem[{{D'Angelo} \& {Spruit}(2011)}]{dangelo2011}
{D'Angelo}, C.~R. \& {Spruit}, H.~C. 2011, \mnras, 416, 893

\bibitem[{{Davidson} \& {Ostriker}(1973)}]{davidson1973}
{Davidson}, K. \& {Ostriker}, J.~P. 1973, \apj, 179, 585

\bibitem[{{Deloye} \& {Bildsten}(2003)}]{deloye2003}
{Deloye}, C.~J. \& {Bildsten}, L. 2003, \apj, 598, 1217

\bibitem[{{Dunkel} {et~al.}(2006){Dunkel}, {Chluba}, \& {Sunyaev}}]{dunkel2006}
{Dunkel}, J., {Chluba}, J., \& {Sunyaev}, R.~A. 2006, Astronomy Letters, 32,
  257

\bibitem[{{Eggleton}(1983)}]{eggleton1983}
{Eggleton}, P.~P. 1983, \apj, 268, 368

\bibitem[{{Elsner} \& {Lamb}(1977)}]{elsner1977}
{Elsner}, R.~F. \& {Lamb}, F.~K. 1977, \apj, 215, 897

\bibitem[{{Frank} {et~al.}(2002){Frank}, {King}, \& {Raine}}]{frank2002book}
{Frank}, J., {King}, A., \& {Raine}, D.~J. 2002, {Accretion Power in
  Astrophysics: Third Edition}

\bibitem[{{Franklin} \& {Colombo}(1970)}]{franklin1970}
{Franklin}, F.~A. \& {Colombo}, G. 1970, \icarus, 12, 338

\bibitem[{{Galloway} {et~al.}(2002){Galloway}, {Chakrabarty}, {Morgan}, \&
  {Remillard}}]{galloway2002}
{Galloway}, D.~K., {Chakrabarty}, D., {Morgan}, E.~H., \& {Remillard}, R.~A.
  2002, \apjl, 576, L137

\bibitem[{{Hakala} {et~al.}(2011){Hakala}, {Charles}, \& {Muhli}}]{hakala2011}
{Hakala}, P.~J., {Charles}, P.~A., \& {Muhli}, P. 2011, \mnras, 991

\bibitem[{{Homer} {et~al.}(1998){Homer}, {Charles}, \&
  {O'Donoghue}}]{homer1998}
{Homer}, L., {Charles}, P.~A., \& {O'Donoghue}, D. 1998, \mnras, 298, 497

\bibitem[{{Hut} \& {Paczy{\'n}ski}(1984)}]{hut1984}
{Hut}, P. \& {Paczy{\'n}ski}, B. 1984, \apj, 284, 675

\bibitem[{{Illarionov} \& {Sunyaev}(1975)}]{illarionov1975}
{Illarionov}, A.~F. \& {Sunyaev}, R.~A. 1975, \aap, 39, 185

\bibitem[{{in 't Zand} {et~al.}(1998){in 't Zand}, {Heise}, {Muller},
  {Bazzano}, {Cocchi}, {Natalucci}, \& {Ubertini}}]{zand1998}
{in 't Zand}, J.~J.~M., {Heise}, J., {Muller}, J.~M., {et~al.} 1998, \aap, 331,
  L25

\bibitem[{{in't Zand} {et~al.}(2005){in't Zand}, {Cumming}, {van der Sluys},
  {Verbunt}, \& {Pols}}]{zand2005}
{in't Zand}, J.~J.~M., {Cumming}, A., {van der Sluys}, M.~V., {Verbunt}, F., \&
  {Pols}, O.~R. 2005, \aap, 441, 675

\bibitem[{{in't Zand} {et~al.}(2007){in't Zand}, {Jonker}, \&
  {Markwardt}}]{zand2007}
{in't Zand}, J.~J.~M., {Jonker}, P.~G., \& {Markwardt}, C.~B. 2007, \aap, 465,
  953

\bibitem[{{Ivanova} {et~al.}(2005){Ivanova}, {Rasio}, {Lombardi}, {Dooley}, \&
  {Proulx}}]{ivanova2005}
{Ivanova}, N., {Rasio}, F.~A., {Lombardi}, Jr., J.~C., {Dooley}, K.~L., \&
  {Proulx}, Z.~F. 2005, \apjl, 621, L109

\bibitem[{{Jeffrey}(1986)}]{jeffrey1986}
{Jeffrey}, L.~C. 1986, \nat, 319, 384

\bibitem[{{Jonker} {et~al.}(2011){Jonker}, {Bassa}, {Nelemans}, {Steeghs},
  {Torres}, {Maccarone}, {Hynes}, {Greiss}, {Clem}, {Dieball}, {Mikles},
  {Britt}, {Gossen}, {Collazzi}, {Wijnands}, {In't Zand}, {M{\'e}ndez}, {Rea},
  {Kuulkers}, {Ratti}, {van Haaften}, {Heinke}, {{\"O}zel}, {Groot}, \&
  {Verbunt}}]{jonker2011}
{Jonker}, P.~G., {Bassa}, C.~G., {Nelemans}, G., {et~al.} 2011, \apjs, 194, 18

\bibitem[{{Kaaret} {et~al.}(2006){Kaaret}, {Morgan}, {Vanderspek}, \&
  {Tomsick}}]{kaaret2006}
{Kaaret}, P., {Morgan}, E.~H., {Vanderspek}, R., \& {Tomsick}, J.~A. 2006,
  \apj, 638, 963

\bibitem[{{King} \& {Begelman}(1999)}]{king1999}
{King}, A.~R. \& {Begelman}, M.~C. 1999, \apjl, 519, L169

\bibitem[{{Kluzniak} {et~al.}(1988){Kluzniak}, {Ruderman}, {Shaham}, \&
  {Tavani}}]{kluzniak1988}
{Kluzniak}, W., {Ruderman}, M., {Shaham}, J., \& {Tavani}, M. 1988, \nat, 334,
  225

\bibitem[{{Krimm} {et~al.}(2007){Krimm}, {Markwardt}, {Deloye}, {Romano},
  {Chakrabarty}, {Campana}, {Cummings}, {Galloway}, {Gehrels}, {Hartman},
  {Kaaret}, {Morgan}, \& {Tueller}}]{krimm2007}
{Krimm}, H.~A., {Markwardt}, C.~B., {Deloye}, C.~J., {et~al.} 2007, \apjl, 668,
  L147

\bibitem[{{Lamb} {et~al.}(1973){Lamb}, {Pethick}, \& {Pines}}]{lamb1973}
{Lamb}, F.~K., {Pethick}, C.~J., \& {Pines}, D. 1973, \apj, 184, 271

\bibitem[{{Landau} \& {Lifshitz}(1975)}]{landau1975}
{Landau}, L.~D. \& {Lifshitz}, E.~M. 1975, {The classical theory of fields}

\bibitem[{{Lasota}(2001)}]{lasota2001}
{Lasota}, J. 2001, \nar, 45, 449

\bibitem[{{Lasota}(2007)}]{lasota2007}
{Lasota}, J. 2007, Comptes Rendus Physique, 8, 45

\bibitem[{{Lasota}(2008)}]{lasota2008}
{Lasota}, J. 2008, \nar, 51, 752

\bibitem[{{Lattimer} \& {Schutz}(2005)}]{lattimer2005}
{Lattimer}, J.~M. \& {Schutz}, B.~F. 2005, \apj, 629, 979

\bibitem[{{Li} \& {Wang}(1999)}]{li1999}
{Li}, X.-D. \& {Wang}, Z.-R. 1999, \apj, 513, 845

\bibitem[{{Lin} \& {Papaloizou}(1979)}]{lin1979}
{Lin}, D.~N.~C. \& {Papaloizou}, J. 1979, \mnras, 186, 799

\bibitem[{{Liu} {et~al.}(2007){Liu}, {van Paradijs}, \& {van den
  Heuvel}}]{liu2007}
{Liu}, Q.~Z., {van Paradijs}, J., \& {van den Heuvel}, E.~P.~J. 2007, \aap,
  469, 807

\bibitem[{{Lovelace} {et~al.}(1999){Lovelace}, {Romanova}, \&
  {Bisnovatyi-Kogan}}]{lovelace1999}
{Lovelace}, R.~V.~E., {Romanova}, M.~M., \& {Bisnovatyi-Kogan}, G.~S. 1999,
  \apj, 514, 368

\bibitem[{{Ma} \& {Li}(2009)}]{ma2009}
{Ma}, B. \& {Li}, X.-D. 2009, \apj, 698, 1907

\bibitem[{{Markwardt} {et~al.}(2003{\natexlab{a}}){Markwardt}, {Juda}, \&
  {Swank}}]{markwardt2003b}
{Markwardt}, C.~B., {Juda}, M., \& {Swank}, J.~H. 2003{\natexlab{a}}, ATel,
  127, 1

\bibitem[{{Markwardt} {et~al.}(2003{\natexlab{b}}){Markwardt}, {Smith}, \&
  {Swank}}]{markwardt2003}
{Markwardt}, C.~B., {Smith}, E., \& {Swank}, J.~H. 2003{\natexlab{b}}, ATel,
  122, 1

\bibitem[{{Markwardt} {et~al.}(2002){Markwardt}, {Swank}, {Strohmayer}, {in 't
  Zand}, \& {Marshall}}]{markwardt2002}
{Markwardt}, C.~B., {Swank}, J.~H., {Strohmayer}, T.~E., {in 't Zand},
  J.~J.~M., \& {Marshall}, F.~E. 2002, \apjl, 575, L21

\bibitem[{{Marsh} {et~al.}(2004){Marsh}, {Nelemans}, \& {Steeghs}}]{marsh2004}
{Marsh}, T.~R., {Nelemans}, G., \& {Steeghs}, D. 2004, \mnras, 350, 113

\bibitem[{{Menou} {et~al.}(2002){Menou}, {Perna}, \& {Hernquist}}]{menou2002}
{Menou}, K., {Perna}, R., \& {Hernquist}, L. 2002, \apjl, 564, L81

\bibitem[{{Morris} {et~al.}(1990){Morris}, {Liebert}, {Stocke}, {Gioia},
  {Schild}, \& {Wolter}}]{morris1990}
{Morris}, S.~L., {Liebert}, J., {Stocke}, J.~T., {et~al.} 1990, \apj, 365, 686

\bibitem[{{Narayan} \& {Yi}(1994)}]{narayan1994}
{Narayan}, R. \& {Yi}, I. 1994, \apjl, 428, L13

\bibitem[{{Nelemans}(2009)}]{nelemans2009}
{Nelemans}, G. 2009, Classical and Quantum Gravity, 26, 094030

\bibitem[{{Nelemans} \& {Jonker}(2010)}]{nelemans2010b}
{Nelemans}, G. \& {Jonker}, P.~G. 2010, \nar, 54, 87

\bibitem[{{Nelemans} {et~al.}(2004){Nelemans}, {Jonker}, {Marsh}, \& {van der
  Klis}}]{nelemans2004}
{Nelemans}, G., {Jonker}, P.~G., {Marsh}, T.~R., \& {van der Klis}, M. 2004,
  \mnras, 348, L7

\bibitem[{{Nelemans} {et~al.}(2006){Nelemans}, {Jonker}, \&
  {Steeghs}}]{nelemans2006}
{Nelemans}, G., {Jonker}, P.~G., \& {Steeghs}, D. 2006, \mnras, 370, 255

\bibitem[{{Nelson} \& {Rappaport}(2003)}]{nelson2003}
{Nelson}, L.~A. \& {Rappaport}, S. 2003, \apj, 598, 431

\bibitem[{{Nelson} {et~al.}(1986){Nelson}, {Rappaport}, \& {Joss}}]{nelson1986}
{Nelson}, L.~A., {Rappaport}, S.~A., \& {Joss}, P.~C. 1986, \apj, 304, 231

\bibitem[{{Osaki}(1974)}]{osaki1974}
{Osaki}, Y. 1974, \pasj, 26, 429

\bibitem[{{Osaki}(1989)}]{osaki1989}
{Osaki}, Y. 1989, \pasj, 41, 1005

\bibitem[{{Paczy{\'n}ski}(1971)}]{paczynski1971}
{Paczy{\'n}ski}, B. 1971, \araa, 9, 183

\bibitem[{{Paczy{\'n}ski}(1977)}]{paczynski1977}
{Paczy{\'n}ski}, B. 1977, \apj, 216, 822

\bibitem[{{Papaloizou} \& {Pringle}(1977)}]{papaloizou1977}
{Papaloizou}, J. \& {Pringle}, J.~E. 1977, \mnras, 181, 441

\bibitem[{{Podsiadlowski} {et~al.}(2002){Podsiadlowski}, {Rappaport}, \&
  {Pfahl}}]{podsiadlowski2002}
{Podsiadlowski}, P., {Rappaport}, S., \& {Pfahl}, E.~D. 2002, \apj, 565, 1107

\bibitem[{{Pols} \& {Marinus}(1994)}]{pols1994}
{Pols}, O.~R. \& {Marinus}, M. 1994, \aap, 288, 475

\bibitem[{{Priedhorsky} \& {Verbunt}(1988)}]{priedhorsky1988}
{Priedhorsky}, W.~C. \& {Verbunt}, F. 1988, \apj, 333, 895

\bibitem[{{Rappaport} {et~al.}(1987){Rappaport}, {Ma}, {Joss}, \&
  {Nelson}}]{rappaport1987}
{Rappaport}, S., {Ma}, C.~P., {Joss}, P.~C., \& {Nelson}, L.~A. 1987, \apj,
  322, 842

\bibitem[{{Rappaport} {et~al.}(2004){Rappaport}, {Fregeau}, \&
  {Spruit}}]{rappaport2004}
{Rappaport}, S.~A., {Fregeau}, J.~M., \& {Spruit}, H. 2004, \apj, 606, 436

\bibitem[{{Ruderman} {et~al.}(1989){Ruderman}, {Shaham}, \&
  {Tavani}}]{ruderman1989}
{Ruderman}, M., {Shaham}, J., \& {Tavani}, M. 1989, \apj, 336, 507

\bibitem[{{Ruderman} \& {Shaham}(1983)}]{ruderman1983}
{Ruderman}, M.~A. \& {Shaham}, J. 1983, \nat, 304, 425

\bibitem[{{Ruderman} \& {Shaham}(1985)}]{ruderman1985}
{Ruderman}, M.~A. \& {Shaham}, J. 1985, \apj, 289, 244

\bibitem[{{Savonije} {et~al.}(1986){Savonije}, {de Kool}, \& {van den
  Heuvel}}]{savonije1986}
{Savonije}, G.~J., {de Kool}, M., \& {van den Heuvel}, E.~P.~J. 1986, \aap,
  155, 51

\bibitem[{{Schreiber} \& {Lasota}(2007)}]{schreiber2007}
{Schreiber}, M.~R. \& {Lasota}, J.-P. 2007, \aap, 473, 897

\bibitem[{{Schulz} {et~al.}(2001){Schulz}, {Chakrabarty}, {Marshall},
  {Canizares}, {Lee}, \& {Houck}}]{schulz2001}
{Schulz}, N.~S., {Chakrabarty}, D., {Marshall}, H.~L., {et~al.} 2001, \apj,
  563, 941

\bibitem[{{Shaham}(1992)}]{shaham1992}
{Shaham}, J. 1992, in X-Ray Binaries and the Formation of Binary and
  Millisecond Radio Pulsars, 375--386

\bibitem[{{Shahbaz} {et~al.}(2008){Shahbaz}, {Watson}, {Zurita}, {Villaver}, \&
  {Hernandez-Peralta}}]{shahbaz2008}
{Shahbaz}, T., {Watson}, C.~A., {Zurita}, C., {Villaver}, E., \&
  {Hernandez-Peralta}, H. 2008, \pasp, 120, 848

\bibitem[{{Shibazaki} {et~al.}(1989){Shibazaki}, {Murakami}, {Shaham}, \&
  {Nomoto}}]{shibazaki1989}
{Shibazaki}, N., {Murakami}, T., {Shaham}, J., \& {Nomoto}, K. 1989, \nat, 342,
  656

\bibitem[{{Simpson} \& {Wood}(1998)}]{simpson1998}
{Simpson}, J.~C. \& {Wood}, M.~A. 1998, \apj, 506, 360

\bibitem[{{Soberman} {et~al.}(1997){Soberman}, {Phinney}, \& {van den
  Heuvel}}]{soberman1997}
{Soberman}, G.~E., {Phinney}, E.~S., \& {van den Heuvel}, E.~P.~J. 1997, \aap,
  327, 620

\bibitem[{{Spruit} \& {Taam}(1993)}]{spruit1993}
{Spruit}, H.~C. \& {Taam}, R.~E. 1993, \apj, 402, 593

\bibitem[{{Strohmayer} {et~al.}(2008){Strohmayer}, {Markwardt}, \&
  {Kuulkers}}]{strohmayer2008}
{Strohmayer}, T.~E., {Markwardt}, C.~B., \& {Kuulkers}, E. 2008, \apjl, 672,
  L37

\bibitem[{{Tauris} \& {Savonije}(1999)}]{tauris1999}
{Tauris}, T.~M. \& {Savonije}, G.~J. 1999, \aap, 350, 928

\bibitem[{{Tsugawa} \& {Osaki}(1997)}]{tsugawa1997}
{Tsugawa}, M. \& {Osaki}, Y. 1997, \pasj, 49, 75

\bibitem[{{Tutukov} \& {Yungelson}(1993)}]{tutukov1993}
{Tutukov}, A.~V. \& {Yungelson}, L.~R. 1993, Astronomy Reports, 37, 411

\bibitem[{{Ustyugova} {et~al.}(2006){Ustyugova}, {Koldoba}, {Romanova}, \&
  {Lovelace}}]{ustyugova2006}
{Ustyugova}, G.~V., {Koldoba}, A.~V., {Romanova}, M.~M., \& {Lovelace},
  R.~V.~E. 2006, \apj, 646, 304

\bibitem[{{van den Heuvel}(1984)}]{vandenheuvel1984}
{van den Heuvel}, E.~P.~J. 1984, Journal of Astrophysics and Astronomy, 5, 209

\bibitem[{{van den Heuvel} \& {Bonsema}(1984)}]{vandenheuvel1984bons}
{van den Heuvel}, E.~P.~J. \& {Bonsema}, P.~F.~J. 1984, \aap, 139, L16

\bibitem[{{van der Sluys} {et~al.}(2005){van der Sluys}, {Verbunt}, \&
  {Pols}}]{sluys2005a}
{van der Sluys}, M.~V., {Verbunt}, F., \& {Pols}, O.~R. 2005, \aap, 431, 647

\bibitem[{{Verbunt}(1987)}]{verbunt1987}
{Verbunt}, F. 1987, \apjl, 312, L23

\bibitem[{{Verbunt} \& {Rappaport}(1988)}]{verbunt1988}
{Verbunt}, F. \& {Rappaport}, S. 1988, \apj, 332, 193

\bibitem[{{Voss} \& {Gilfanov}(2007)}]{voss2007}
{Voss}, R. \& {Gilfanov}, M. 2007, \mnras, 380, 1685

\bibitem[{{Wang} {et~al.}(2011){Wang}, {Zhang}, {Zhao}, {Kojima}, {Yin}, \&
  {Song}}]{wang2011}
{Wang}, J., {Zhang}, C.~M., {Zhao}, Y.~H., {et~al.} 2011, \aap, 526, A88+

\bibitem[{{Warner}(2003)}]{warner2003book}
{Warner}, B. 2003, {Cataclysmic Variable Stars}, ed. {Warner, B.}

\bibitem[{{Webbink}(1985)}]{webbink1985}
{Webbink}, R.~F. 1985, {Stellar evolution and binaries}, ed. {Pringle, J.~E.~\&
  Wade, R.~A.}, 39--+

\bibitem[{{Wijnands}(2010)}]{wijnands2010}
{Wijnands}, R. 2010, Highlights of Astronomy, 15, 121

\bibitem[{{Wood} {et~al.}(2009){Wood}, {Thomas}, \& {Simpson}}]{wood2009}
{Wood}, M.~A., {Thomas}, D.~M., \& {Simpson}, J.~C. 2009, \mnras, 398, 2110

\bibitem[{{Yungelson}(2008)}]{yungelson2008}
{Yungelson}, L.~R. 2008, Astronomy Letters, 34, 620

\bibitem[{{Yungelson} {et~al.}(2006){Yungelson}, {Lasota}, {Nelemans}, {Dubus},
  {van den Heuvel}, {Dewi}, \& {Portegies Zwart}}]{yungelson2006}
{Yungelson}, L.~R., {Lasota}, J., {Nelemans}, G., {et~al.} 2006, \aap, 454, 559

\bibitem[{{Yungelson} {et~al.}(2002){Yungelson}, {Nelemans}, \& {van den
  Heuvel}}]{yungelson2002}
{Yungelson}, L.~R., {Nelemans}, G., \& {van den Heuvel}, E.~P.~J. 2002, \aap,
  388, 546

\bibitem[{{Zhang} \& {Kojima}(2006)}]{zhang2006}
{Zhang}, C.~M. \& {Kojima}, Y. 2006, \mnras, 366, 137

\end{thebibliography}

\appendix

\section{Fitted tracks}
\label{wdfits}

Sometimes it is useful to have simple analytic estimates of the evolution of the quantities involved. For this purpose we provide powerlaw approximations of the tracks with full feedback of angular momentum, optimized for system age (time since the onset of mass transfer) $t \in [1\ \mbox{Myr}, 20\ \mbox{Gyr}]$,

\begin{eqnarray}
    \label{eqpara}
    \dot{M}_\mathrm{d} &\approx& f_{1} \cdot t^{e_{1}} \\
    P_\mathrm{orb} &\approx& f_{2} \cdot t^{e_{2}} \nonumber \\
    M_\mathrm{d} &\approx& f_{3} \cdot t^{e_{3}} \nonumber \\
    \dot{M}_\mathrm{d} &\approx& f_{4} \cdot P_\mathrm{orb}^{e_{4}} \nonumber \\
    M_\mathrm{d} &\approx& f_{5} \cdot P_\mathrm{orb}^{e_{5}} \nonumber \\
    \dot{M}_\mathrm{d} &\approx& f_{6} \cdot M_\mathrm{d}^{e_{6}} \nonumber
\end{eqnarray}
with $t$ in \mbox{yr}, the mass transfer rate $\dot{M}_\mathrm{d}$ in $M_{\odot} \mbox{yr}^{-1}$, the orbital period $P_\mathrm{orb}$ in \mbox{min} and the donor mass $M_\mathrm{d}$ in $M_{\odot}$. The corresponding parameters $f_\mathrm{i}$ and $e_\mathrm{i}$ for accretor masses $M_\mathrm{a} = 1.4\ M_{\odot}$ and $10\ M_{\odot}$ are listed in Table \ref{table:para}. The exponent in the $\dot{M}_\mathrm{d}(t)$ relation ($e_{1}$) must be roughly $1$ lower than in the $M_\mathrm{d}(t)$ relation ($e_{3}$), but not necessarily precisely $1$ because both tracks have been fitted separately.

For an initially $1.4\ M_{\odot}$ neutron star accretor, these relations fit the modeled helium white dwarf tracks (shown in Figs. \ref{fig:wdtracks}, \ref{fig:wddonor} and \ref{fig:wdtime}) with an average error of $4.3\%$, in the tracks $t = 1\ \mbox{Myr}$ corresponds to $P_\mathrm{orb}=9.0\ \mbox{min}$, $\dot{M}_\mathrm{d}=2.1 \cdot 10^{-8}\ M_{\odot} \mbox{yr}^{-1}$ (lower than the Eddington limit; mass transfer is conservative for the whole range) and $M_\mathrm{d} = 0.074\ M_{\odot}$.

Similarly for an initially $10\ M_{\odot}$ black hole accretor, where the average error is $5.3\%$.
For a $10\ M_{\odot}$ black hole, $t = 1\ \mbox{Myr}$ corresponds to $P_\mathrm{orb}=12.3\ \mbox{min}$, $\dot{M}_\mathrm{d}=1.4 \cdot 10^{-8}\ M_{\odot} \mbox{yr}^{-1}$ and $M_\mathrm{d} = 0.050\ M_{\odot}$.

From the fitted parameters as well as in Fig. \ref{fig:wdtracks}, it can be seen that all slopes are nearly independent of accretor mass. Also, the $M_\mathrm{d}(P_\mathrm{orb})$ relations are rather similar for both accretor masses, as expected.

\begin{table}
\caption{Values for the parameters in Eqs. (\ref{eqpara}).}
\label{table:para}
\begin{tabular}{|l|ll|ll|}
\hline\hline
 & \multicolumn{2}{l|}{Fitted tracks} & \multicolumn{2}{l|}{Analytic approximations}\\
\hline
Parameter & $1.4\ M_{\odot}$ & $10\ M_{\odot}$ & $1.4\ M_{\odot}$ & $10\ M_{\odot}$ \\
\hline
$f_{1}$ & $1.18$ & $0.882$ & $0.736$ & $0.515$ \\
$f_{2}$ & $0.326$ & $0.482$ & $0.196$ & $0.281$ \\
$f_{3}$ & $5.92$ & $4.62$ & $2.70$ & $1.89$ \\
$f_{4}$ & $3.01 \cdot 10^{-3}$ & $1.61 \cdot 10^{-2}$ & $3.71 \cdot 10^{-4}$ & $1.38 \cdot 10^{-3}$ \\
$f_{5}$ & $1.38$ & $1.69$ & \multicolumn{2}{c|}{$0.530$}  \\
$f_{6}$ & $7.75 \cdot 10^{-4}$ & $1.87 \cdot 10^{-3}$ & $7.15 \cdot 10^{-3}$ & $2.65 \cdot 10^{-2}$ \\
$e_{1}$ & $-1.29$ & $-1.29$ & \multicolumn{2}{c|}{$-14/11\ (\approx -1.27)$} \\
$e_{2}$ & $0.242$ & $0.237$ & \multicolumn{2}{c|}{$3/11\ (\approx 0.273)$} \\
$e_{3}$ & $-0.313$ & $-0.323$ & \multicolumn{2}{c|}{$-3/11\ (\approx -0.273)$} \\
$e_{4}$ & $-5.32$ & $-5.45$ & \multicolumn{2}{c|}{$-14/3\ (\approx -4.67)$} \\
$e_{5}$ & $-1.29$ & $-1.36$ & \multicolumn{2}{c|}{$-1$} \\
$e_{6}$ & $4.11$ & $4.01$ & \multicolumn{2}{c|}{$14/3\ (\approx 4.67)$} \\
\hline
\end{tabular}
\tablefoot{
$1.4\ M_{\odot}$ and $10\ M_{\odot}$ are the accretor masses. The fitted tracks are discussed in Appendix \ref{wdfits} and the analytic approximations in Appendix \ref{appa}. The analytic parameters $e_\mathrm{i}$ as well as $f_{5}$ belong to both accretor masses.
}
\end{table}

\section{Approximate analytic solution}
\label{appa}

By making a few mild assumptions it is possible to solve almost the entire evolution (assuming full feedback of angular momentum to the orbit) of a binary containing a zero-temperature white dwarf donor analytically, except for the early stages when the donor mass is high. These derivations only serve to gain physical insight into the evolution of UCXBs -- throughout the paper numerical solutions of more precise equations have been used.

First we approximate the low-mass white dwarf mass-radius relation from Eq. (\ref{rwd}) by the non-relativistic relation

\begin{equation}
    \label{nonrel}
    R_\mathrm{d}/R_{\odot} = 10^{-2} (M_\mathrm{d}/M_{\odot})^{-1/3},
\end{equation}
which is accurate around $M_\mathrm{d} = 0.01\ M_{\odot}$ (and within $20\%$ of the radius given by Eq. (\ref{rwd}) for $4 \cdot 10^{-3} < M_\mathrm{d}/M_{\odot} < 1$), but still represents the largest simplification made here. By substituting the white dwarf radius from Eq. (\ref{nonrel}) into the Roche-lobe radius in the well-known period-mean density relation $P_\mathrm{orb} = 9\pi \sqrt{R_\mathrm{L}^{3}/(2GM_\mathrm{d})}$ we find the orbital period of a system with a white dwarf donor,

\begin{equation}
    \label{pnonrel}
    P_\mathrm{orb} = A M_\mathrm{d}^{-1}
\end{equation}
with the constant

\begin{equation}
    A = 9\pi 10^{-3} \sqrt{\frac{M_{\odot} R_{\odot}^{3}}{2 G}} \approx 0.53\ M_{\odot}\mbox{min}.
\end{equation}
The orbital period is inversely proportional to donor mass (and proportional to donor volume).
The mass transfer rate is given by Eq. (\ref{mt2}), in which $\zeta_\mathrm{d} = -1/3$ follows from Eq. (\ref{nonrel}) and $\zeta_\mathrm{L} = -5/3$ is the low mass ratio limit of Eq. (\ref{zetarl}) for $\eta = 0$. Furthermore, we can use the gravitational wave equation and Kepler's third law to replace $a$ by $P_\mathrm{orb}$ to obtain

\begin{equation}
    \dot{M}_\mathrm{d} = -\frac{48}{5} \frac{G^{5/3}}{c^{5}} \frac{M_\mathrm{a}}{M_\mathrm{tot}^{1/3}} M_\mathrm{d}^{2} \left( \frac{2\pi}{P_\mathrm{orb}} \right)^{8/3}.
\end{equation}
Lastly, $M_\mathrm{d} \ll M_\mathrm{a}$ implies $M_\mathrm{tot} \approx M_\mathrm{a}$, and we substitute Eq. (\ref{pnonrel}) for $P_\mathrm{orb}$ to express the mass transfer rate in only the component masses

\begin{equation}
    \label{mt_m}
    \dot{M}_\mathrm{d} = -C M_\mathrm{a}^{2/3} M_\mathrm{d}^{14/3}
\end{equation}
with the constant

\begin{equation}
    C = \frac{48}{5} \frac{G^{5/3}}{c^{5}} \left( \frac{2\pi}{A} \right)^{8/3} \approx 0.0057\ M_{\odot}^{-13/3}\mbox{yr}^{-1}.
\end{equation}
Equation (\ref{mt_m}) shows that the mass transfer rate depends strongly on donor mass. $8/3$ out of the $14/3$ power is due to massive donors having a shorter semi-major axis when they fill their Roche lobe, which results in stronger gravitational wave radiation. The direct effect of the high donor mass is slightly smaller with $2$ out of $14/3$. Alternatively, the mass transfer rate can be expressed as a function of accretor mass and orbital period

\begin{equation}
    \label{mt_p}
    \dot{M}_\mathrm{d} = -C M_\mathrm{a}^{2/3} \left( \frac{P_\mathrm{orb}}{A} \right)^{-14/3}.
\end{equation}
By integrating the mass transfer rate given by Eq. (\ref{mt_m}), we obtain the system age $t$ as a function of component masses.

\begin{equation}
    \label{mdotint}
    t = \int_{\infty}^{M_\mathrm{d}} \frac{\mathrm{d}M_\mathrm{d}'}{\dot{M}_\mathrm{d}'} = \frac{3}{11 C} M_\mathrm{a}^{-2/3} M_\mathrm{d}^{-11/3}
\end{equation}
where integrating from $\infty$ is justified by the negligibly short timescale associated with a high donor mass, as evidenced by the $-11/3$ power of the donor mass. The donor mass as a function of time follows,

\begin{equation}
    \label{m_t}
    M_\mathrm{d} = \left( \frac{3}{11 C} \right)^{3/11} M_\mathrm{a}^{-2/11} t^{-3/11}
\end{equation}
and by substituting Eq. (\ref{pnonrel})

\begin{equation}
    \label{p_t}
    P_\mathrm{orb} = A \left( \frac{11 C}{3} \right)^{3/11} M_\mathrm{a}^{2/11} t^{3/11}.
\end{equation}
The evolutionary timescale is $\propto P_\mathrm{orb}^{11/3}$, even steeper than the gravitational wave timescale in a detached binary ($\propto P_\mathrm{orb}^{8/3}$) due to the effect of the donor mass. Finally, differentiating Eq. (\ref{m_t}) to time yields

\begin{equation}
    \label{mt_t}
    \dot{M}_\mathrm{d} = -\left( \frac{3}{11} \right)^{14/11} C^{-3/11} M_\mathrm{a}^{-2/11} t^{-14/11}.
\end{equation}
As mentioned in Sect. \ref{macc} it shows that when comparing for systems with the same age, a less massive accretor corresponds to a higher donor mass and mass transfer rate. The mass transfer timescale of systems of fixed age $t$ is given by $|M_\mathrm{d}/\dot{M}_\mathrm{d}| = \frac{11}{3} t$. This value is $2/3$ times the gravitational wave timescale, and is independent of accretor mass in this approximation. As we saw in Sect. \ref{macc}, however, this timescale is very slightly ($\sim$ 7\% for a factor $7$ in accretor mass) shorter for high-mass accretor systems when we use the more precise mass-radius relation.

Above relations are valid for $M_\mathrm{d} \lesssim 0.2\ M_{\odot}$.
Table \ref{table:para} lists numerical values for the parameters in several equations given above.

\subsection{Luminosity}

By assuming an accretor radius and taking the time-averaged bolometric luminosity of an UCXB as

\begin{equation}
    L = -\frac{G M_\mathrm{a} \dot{M}_\mathrm{d}}{2 R_\mathrm{a}}
\end{equation}
where $\dot{M}_\mathrm{d}$ is still defined as negative, and by using Eq. (\ref{mt_p}), we can express the following quantities in the observables $L$ and $P_\mathrm{orb}$:

\begin{equation}
    \label{obs_ma}
    M_\mathrm{a} = \left( \frac{2 R_\mathrm{a}}{C G} L \right)^{3/5} \left( \frac{P_\mathrm{orb}}{A} \right)^{14/5}
\end{equation}
and

\begin{equation}
    \dot{M}_\mathrm{d} = -C^{3/5} \left( \frac{2 R_\mathrm{a}}{G} L \right)^{2/5} \left( \frac{P_\mathrm{orb}}{A} \right)^{-14/5}.
\end{equation}
By combining Eqs. (\ref{p_t}) and (\ref{obs_ma}) the system age follows

\begin{equation}
    t = \frac{3}{11} C^{-3/5} \left( \frac{2 R_\mathrm{a}}{G} L \right)^{-2/5} \left( \frac{P_\mathrm{orb}}{A} \right)^{9/5}.
\end{equation}

\section{First Lagrangian point location approximations}

The distance between the first Lagrangian point (L1) and binary component 2 is given by

\begin{equation} 
    \label{l1fit9}
    \frac{d(\mathrm{L1},\mathrm{2})}{a} = \left( 1 + \left( 3^{\scriptstyle{\tanh(\ln(q)/11.8)}} q \right)^{\scriptstyle{-0.3295 - 1/(21.3 + 0.559 \ln^{2}(q))}} \right)^{-1}
\end{equation}
where the relative errors in $d(\mathrm{L1},2)$ and $d(\mathrm{L1},1)$ are less than $0.015\%$ for $7 \cdot 10^{-4} < q = M_{2}/M_{1} < 1.5 \cdot 10^{3}$ and less than $0.1\%$ for $10^{-6} < q < 10^{6}$.

The next two expressions are only valid for $q \le 1$. The distance between L1 and the least massive binary component (2), valid within $0.02\%$ (again, this is the relative error in the distance to either component) for $7 \cdot 10^{-5} < q = M_{2}/M_{1} \le 1$, is given by

\begin{equation} 
    \label{l1fit6}
    \frac{d(\mathrm{L1},\mathrm{2})}{a} = \left( \frac{55}{49} + q^{-31/42} \ln{\left(1 + \frac{66}{47} q^{29/72}\right) } \right)^{-1}.
\end{equation}

A simpler but less accurate ($< 0.5\%$) expression, inspired by the form of the Eggleton Roche lobe equation (\ref{roche}), is\footnote{By adjusting the first term inside brackets (the 1) to $0.995$, the maximum error becomes $0.3\%$, but the approximation at $q = 1$ becomes worse and matches less well with its continuation at $q > 1$.}

\begin{equation} 
    \label{l1fit}
    \frac{d(\mathrm{L1},\mathrm{2})}{a} = \left( 1 + \frac{ \ln(1 + q^{1/3}) }{ (q^{2}/3)^{1/3} } \right)^{-1}
\end{equation}
where $0 < q = M_{2}/M_{1} \le 1$.

\citet[chap.~2]{warner2003book} lists three approximations of the L1 location, but those are valid over a much smaller range in mass ratio and have much larger errors than our equations.

\section{Roche lobe potential approximations}

Two simpler but less accurate versions of Eq. (\ref{pot}) are given here. For a binary with $0 < q = M_{2}/M_{1} \le 1$, the potential at the Roche lobe surface $\Phi_\mathrm{L}$ is approximated to within $0.4\%$ by

\begin{equation} 
    \label{potfit}
    \frac{-\Phi_\mathrm{L}}{G M_\mathrm{1}/a} = \frac{3/2 + (10 q^{2})^{1/3}}{1 - (q/5)^{3/2}}
\end{equation}
and to within $1\%$ by

\begin{equation} 
    \label{potfit2}
    \frac{-\Phi_\mathrm{L}}{G M_\mathrm{1}/a} = \frac{3}{2} + 2.46 q^{0.73}.
\end{equation}

\end{document}